\documentclass[usenatbib]{mn2e}

\usepackage{graphicx,graphics}
\usepackage{color,longtable,lscape}

\def\aj{AJ}%
\def\araa{ARA\&A}%
\def\apj{ApJ}%
\def\apjl{ApJ}%
\def\apjs{ApJS}%
\def\aap{A\&A}%
\def\mnras{MNRAS}%
\def\pasp{PASP}%
\def\nat{Nature}%
\def\bain{Bull.~Astron.~Inst.~Netherlands}%
\newcommand{\msun}{M_\odot}
\newcommand{\rhm}{r_{\rm hm}}
\newcommand{\rvir}{r_{\rm vir}}
\newcommand{\kms}{km\,s$^{-1}$}
\newcommand{\trelax}{t_{\rm hm}}

\newcommand{\nsb}{N_{\rm sb}}
\newcommand{\ns}{N_{\rm s}}
\newcommand{\nb}{N_{\rm b}}
\newcommand{\np}{N_{\rm p}}
\newcommand{\ratio}{\mathcal{R}}
\newcommand{\ratiofit}{\mathcal{R}_f}
\newcommand{\alphafit}{\alpha_f}
\newcommand{\binfrac}{\mathcal{B}}

\newcommand{\mstar}{M_{\rm s}}
\newcommand{\mparticle}{M_{\rm sb}}
\newcommand{\velocitytype}{f_{\rm ej}}

\newcommand{\thalfstar}{t_{\rm hs}}
\newcommand{\thalfplanet}{t_{\rm hp}}
\newcommand{\tdiss}{t_{\rm diss}}

\newcommand{\periastron}{p}
\newcommand{\vinf}{v_\infty}
\newcommand{\vperiastron}{v_p}
\newcommand{\vpoffset}{v_o}
\newcommand{\vpmax}{v_m}

\newcommand{\nenc}{\mathcal{N}}

\title[Free-floating planets in star clusters]{Close encounters involving free-floating planets in star clusters}  
\author[Wang, Kouwenhoven, Zheng, Church \& Davies]{Long Wang$^{1,2}$\thanks{E-mail: long.wang@pku.edu.cn}, M.~B.~N. Kouwenhoven$^{2,1}$, Xiaochen Zheng$^{1,2}$, 
\newauthor{ Ross~P.~Church$^{3}$, \& Melvyn~B.~Davies$^{3}$}\\
  $^{1}$Department of Astronomy, Peking
  University, Yi He Yuan Lu 5, Haidian Qu, Beijing 100871, P.R.~China\\
  $^{2}$Kavli Institute for Astronomy and Astrophysics, Peking
  University, Yi He Yuan Lu 5, Haidian Qu, Beijing 100871, P.R.~China\\
  $^{3}$Department of Astronomy and Theoretical
  Physics, Lund Observatory, Box 43, SE-221 00, Lund, Sweden}

\begin{document}

\date{Accepted --. Received --; in original form --}

\pagerange{---} \pubyear{---}

\maketitle

\label{firstpage}

\begin{abstract}
Instabilities in planetary systems can result in the ejection of planets from their host system, resulting in free-floating planets (FFPs). If this occurs in a star cluster, the FFP may remain bound to the star cluster for some time and interact with the other cluster members until it is ejected. Here, we use $N$-body simulations to characterise close star-planet and planet-planet encounters and the dynamical fate of the FFP population in star clusters containing $500-2000$ single or binary star members. We find that FFPs ejected from their planetary system at low velocities typically leave the star cluster 40\% earlier than their host stars, and experience tens of close ($<1000$~AU) encounters with other stars and planets before they escape. 
The fraction of FFPs that experiences a close encounter depends on both the stellar density and the initial velocity distribution of the FFPs. Approximately half of the close encounters occur within the first 30~Myr, and only 10\% occur after 100~Myr. The periastron velocity distribution for all encounters is well-described by a modified Maxwell-Bolzmann distribution, and the periastron distance distribution is linear over almost the entire range of distances considered, and flattens off for very close encounters due to strong gravitational focusing. Close encounters with FFPs can perturb existing planetary systems and their debris structures, and they can result in re-capture of FFPs. In addition, these FFP populations may be observed in young star clusters in imaging surveys; a comparison between observations and dynamical predictions may provide clues to the early phases of stellar and planetary dynamics in star clusters.
\end{abstract}

\begin{keywords}
Open clusters and associations: general; stars: kinematics and dynamics; planets: dynamical evolution and stability
\end{keywords}

\section{Introduction}

The majority of stars form in clustered environments \citep[e.g.,][]{lada2003, portegies2010}, where close encounters between stars are frequent. These close encounters can perturb or destroy planetary systems \citep[e.g.,][]{spurzem2009, boley2012}. Even a mild perturbation can break up a marginally stable planetary system and can even affect the shortest-period planets in the system \citep[e.g.,][]{hao2013}, which can result in strong planet-planet scattering, physical collisions between planets and their host stars or other planets \citep[e.g.,][]{Rasio1996, Chatterjee2008, Juric2008, Nagasawa2011}, long-term secular evolution \citep[e.g.,][]{malmberg2007_kozai, parker2009_kozai, malmberg2011}, and the ejection of planets from the system. In addition, mass loss due to stellar evolution of single or binary host stars can result in loss of planetary companions through similar interactions \citep[e.g.,][]{veras2011, veras2012, adams2013, voyatzis2013, nowak2013}. Even the Galactic tidal field can indirectly play a role in the disruption of planetary systems when a wide stellar companion is present \citep[e.g.,][]{kaib2013}. \cite{verasscattering2012} demonstrate that the dynamical and stellar evolution of isolated planetary systems alone cannot account for the observed free-floating planet (FFP) population in the Galactic field, and that close encounters in star clusters are an important source of FFPs in the field.

Although only a small number of FFP candidates have been detected so far, they are potentially abundant in the Galactic disk \cite[e.g.,][]{strigari2012}. Exoplanets that are too distant from stars to directly affect their observable properties are most easily detected using microlensing \cite[e.g.,][]{mao1991, gould1992, abe2004, beaulieu2006, gaudi2012}. Microlensing surveys also have the potential to discover FFPs \citep[e.g.,][]{distefano2012},
and results imply that there are roughly twice as many FFPs in the Solar neighbourhood than there are main-sequence stars \citep{sumi2011}. In addition, deep imaging surveys can be used to detect young planetary-mass objects near the deuterium burning limit in young star clusters \citep[e.g.,][]{lucas2006, caballero2007, bihain2009, quanz2010, penaramirez2011, scholz2012, delorme2012}. These observational studies help us further constrain the origin and fate of these FFPs provided that a good understanding of their dynamics is known.

FFPs are thought to have been ejected from their system with velocities of typically $0.1-10$~\kms, as a result of planet-planet scattering ({\em delayed ejection}) or immediately after a close encounter with a passing star ({\em prompt ejection}), and the ejection velocities from the latter process tend to be higher \citep{malmberg2011}. When a FFP is ejected in a star cluster, it may escape immediately if its ejection velocity exceeds the local escape velocity of the star cluster, or it may remain bound to its host cluster for millions of years, until it escapes through ejection or evaporation. During its life in a star cluster, a FFP can experience multiple close encounters with other stars before escaping, and may even be re-captured by another star \citep{kouwenhoven2010, malmberg2011, moeckel2011, parker2012, peretskouwenhoven}. 
Direct $N$-body simulations of single-planet systems in star clusters have shown that many are disrupted and that the resulting FFPs can remain in these star clusters for many millions of years \citep[e.g.,][]{hurley2002, parker2012, craig2013}. Fly-by simulations mimicking the evolution of multi-planet systems in star clusters confirm that the survival of these systems depends strongly on the properties of the stellar environment and the semi-major axes of the planets, but also demonstrated that planetary multiplicity itself plays an important roles as planets in perturbed systems also mutually interact \citep[e.g.,][]{Chatterjee2012, hao2013, liu2013}.

The aim of this study is to analyse the dynamical properties of FFPs in low-mass star clusters, with a particular focus on close encounters between the members of the star cluster (single stars, binary stars, and FFPs). This article is organised as follows. The methods and assumptions are described in \S~\ref{section:method}. The results are presented in \S~\ref{section:results}. Finally, we draw the conclusions and describe our future work in \S~\ref{section:conclusions}.


\section{Method and assumptions} \label{section:method}


\subsection{Initial conditions for the star clusters}

\begin{table}
  \caption{Initial conditions for the modelled star clusters. \label{table:starcluster} }
  \begin{tabular}{ll}
    \hline\hline
    Quantity               & Value \\
    \hline
    Number of stars \& binaries & $N=\ns+\nb = 500$, 1000, 2000 \\
    Half-mass radius & $\rhm = 0.38$, 0.77, 1.54~pc\\
    Dynamical model        & \cite{plummer1911} \\
    Virial ratio           & $Q=1/2$ \\
    Tidal field            & Galactic Solar orbit \\
    \hline
    Initial mass function  & \cite{kroupatoutgilmore}, $0.2-5\msun$ \\
    Binary fraction        & $\binfrac=\nb/N = 0\%$, $20\%$, $50\%$ \\ 
    Semi-major axis distr. & $f_a(a) \propto a^{-1}$; ($10^{-6}-10^{-3})\times\rvir$\\
    Mass ratio distr.      & Random pairing \\
    Eccentricity distr.    & $f_e(e) = 2e$ ($0\leq e<1$) \\
    Orbital orientation    & Random \\
    \hline
    Planet-to-star ratio   & $\ratio=\np/\nsb=0.5, 1, 2$ \\
    Planet mass            & $M_p = 1 M_J$ \\
    Density distribution   & Equivalent to that of stars\\
    Planet ejection velocity & $\mstar$, $\velocitytype$ (see Figure~\ref{figure:ejectionvelocities} and \S~\ref{section:planetinitial}) \\
    \hline\hline
  \end{tabular}
\end{table}

The properties of our model star clusters are summarised in
Table~\ref{table:starcluster}. We create star clusters with
\cite{plummer1911} density and velocity distributions. We carry out simulations of open star clusters with $N=\ns+\nb=500$, 1000 and 2000 systems, where $\ns$ and $\nb$ represent the number of single stars and
binary systems in the star cluster, respectively, and $\nsb=\ns+2\nb$ is the total
number of stars in the cluster.  The initial values for the virial radii are $\rvir=0.5$, 1.0, and 2.0~pc, which are typical for open clusters \citep[e.g.,][]{lada2003}, and correspond to intrinsic half-mass radii of $\rhm = 0.38$, 0.77 and 1.54 pc, since for the Plummer model $\rvir\approx 1.30\,\rhm$ \citep[e.g.,][]{heggiehut}. Each star cluster is initially in
virial equilibrium, i.e., $Q=|K/P|=1/2$, where $K$ and $P$ are the total kinetic
and potential energies, respectively. As a consequence, the corresponding initial (one-dimensional)
velocity dispersion $\sigma(r)$ as a function of distance to the cluster centre $r$
is 
\begin{equation} \label{eq:velocitydispersion}
  \sigma^2(r) = \frac{GM}{6\rhm} \left( 1+ \frac{r^2}{\rhm^2} \right)^{-1/2} \ ,
\end{equation}
where $G$ the gravitational constant
and $M$ the total cluster mass \citep{heggiehut}. 

Following \cite{malmberg2007_encounters}, the stellar masses are drawn from the \cite{kroupatoutgilmore} initial mass function (IMF) in the mass range $0.2-5~\msun$. There is evidence that the upper mass limit may depend on the mass of the star cluster \citep[e.g.,][]{weidner2004, weidner2013}, this dependence is still under debate \citep[e.g.,][]{cervino2013a, cervino2013b}. We therefore follow \cite{malmberg2007_encounters} in choosing a constant upper mass limit, while realising the possibility that realistic clusters may host a more massive star that can alter the dynamics of the FFP population.

 We carry out the simulations with binary fractions $\binfrac=\nb/N=\nb/(\ns+\nb)=0\%$, $20\%$, and $50\%$. The individual components of the binary systems are
randomly paired from the IMF \citep[see][for details]{kouwenhoven2006, kouwenhovenpairing}. For this method of pairing stars into binary systems the total mass of the star cluster equals $N\langle M \rangle (1+\binfrac)$, where $\langle M \rangle$ is the average stellar mass. The
adopted semi-major axis distribution is $f_a(a)\propto a^{-1}$, which corresponds to a flat distribution in $\log a$, also known as
\"{O}pik's law \citep[e.g.,][]{vanalbada1968,vereshchagin1987,poveda2004, kouwenhoven2005, kouwenhoven2007}. Semi-major axes are drawn from this distribution in the range
$(10^{-6}-10^{-3})\times\rvir$, which corresponds to $1-1000$~AU for the star clusters with $\rhm=0.38$~pc. We do
not include very tight binary systems as they are relatively inert and effectively act as
single stars. We also do not include very wide binary systems, as their individual components experience encounters very similar to those of single stars. Moreover, many of these wide binary systems are easily destroyed \citep[e.g.,][]{parker2009_binaries}. We choose a thermal eccentricity distribution $f_e(e)=2e$ for $0 \leq e < 1$ \citep{heggie1975}, and all orbits are assigned a random spatial orientation and a random orbital phase.

We include an external tidal field by modelling the star clusters on a Galactic circular orbit in the solar neighbourhood. The tidal radii of the star clusters are then given by
\begin{equation} \label{eq:tidalfield}
  r_t(M) = \left[ \frac{GM}{4A(A-B)} \right]^{1/3} \approx 1.41 \left(\frac{M}{\msun}\right)^{1/3} \ {\rm pc} \ ,
\end{equation}
where $A=14.4$\,\kms kpc$^{-1}$ and $B=-12.0$\,\kms kpc$^{-1}$ are the Oort
constants \citep[e.g.,][]{binneytremaine}, and $M$ is the total mass of the cluster.  This external tidal field enhances the dissolution of the cluster, in particular for low-mass and low-density star clusters. In our grid of models the tidal radii are $r_t=6.0-11.5$~pc, depending on total mass of the star cluster.


\subsection{Initial conditions for the free-floating planets} \label{section:planetinitial}

\begin{figure}
  \centering
  \includegraphics[width=0.45\textwidth,height=!]{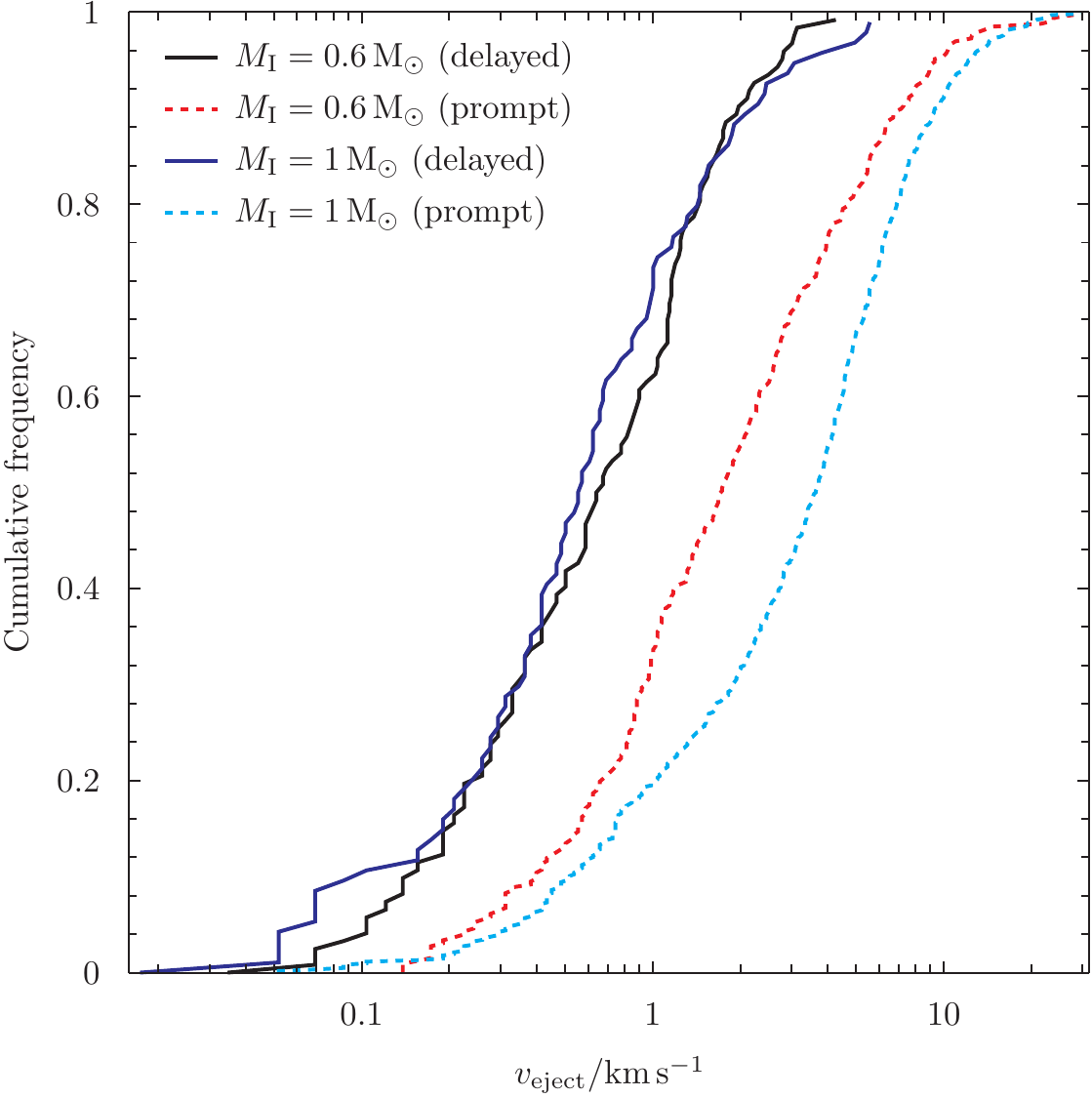}
  \caption{The cumulative distribution of velocities at which planets escape from their host
    stars, as obtained by \protect\cite{malmberg2011}. The four curves indicate the
    results for models with different host star masses, and represent the
    distributions for prompt ejections (during the encounter) and delayed
    ejections (after the encounter). For comparison, we also model the evolution of FFP populations with zero ejection velocities, in addition to the four velocity distributions shown here.
  \label{figure:ejectionvelocities} }
\end{figure}

In addition to the stellar population we also add a population of FFPs. The
initial positions of these planets are drawn from the same distribution as that of
the stars. Since FFPs are thought to have been ejected from planetary systems,
they have an initial position distribution which is statistically identical to
that of their host stars. It can be argued that planets are preferably ejected
from their host systems near the cluster centre, as the close encounter frequency
with other stars is significantly larger in this region. On the other hand, from
delayed ejection in perturbed multi-planet systems, we can expect an initial
spatial distribution of ejected planets that is similar to that of the stars in
the star cluster when the delay time is larger than the crossing time, which is $1-3$~Myr for our cluster models. Fly-by experiments by \cite{malmberg2011}, \cite{hao2013} and \cite{liu2013} have demonstrated that, although some close encounters result in prompt ejections, perturbed planetary systems can eject planets up to tens of millions of years after a close encounter has taken place, at a location that is unrelated to the place where the close encounter took place.

Moreover, we make the assumption that all FFPs are free-floating at the start of our simulations, and that none of the stars have planetary companions. Hence, we ignore the possibility of new FFPs being ejected from perturbed planetary systems in the star clusters. Although this may seem a strong constraint to the applicability of our study, we show in Section~\ref{section:results} that many of our results are also applicable to populations of FFPs that are ejected at different times. Moreover, our results can also be use to calculate encounter probabilities for individual FFPs, and to estimate the time at which an individual FFPs may escape.

Two dynamical limits can be considered here: (i) the limit where all planets have zero ejection velocity with respect to their host stars, and (ii) the limit where all planets have ejection velocities that are substantially larger than the star cluster's escape velocity. The latter case is trivial and results in the immediate ejection of all FFPs. In the former case, both the stellar and planetary populations initially have identical position and velocity distributions. Subsequently, energy exchange between the bodies, result in mass segregation and also in the gradual ejection of all planets from the system.

The adopted ejection velocity distributions $\velocitytype$ are based on the work of \cite{malmberg2011}, who studied the evolution of planetary systems consisting of four gas-giant planets orbiting a star in a star cluster. They studied the prompt ejection
velocities of planets escaping during close encounters with other stars ($\velocitytype=P$), and the delayed ejections as a result of planet-planet scattering in a perturbed system ($\velocitytype=D$). Their simulations were carried out using host stars with masses of $\mstar=0.6\msun$ and $\mstar=1.0\msun$. The
cumulative ejection velocities for the different models are shown in
Figure~\ref{figure:ejectionvelocities}. For comparison, we also carry out
simulations with zero ejection velocities, $\velocitytype=Z$, where the initial velocity
distributions of planets and stars are identical. We study the dynamical evolution
of the FFP population for each of these five distributions $\velocitytype$. As the two delayed ejection velocity distributions in Figure~\ref{figure:ejectionvelocities} are similar, the results for the models with $\mstar=1.0\msun$ and $\velocitytype=D$ are omitted in most figures and discussion throughout the remaining part of this paper.

Each planet is assigned a mass of $M_P=1~M_J \approx 10^{-3}\msun$. Since these planets are of very low mass compared to the stars, the actual value of
the mass does not affect the evolution of the stellar population and star
cluster as a whole, the star-planet encounter rate, or the rate of planet
capture by stars. It may, however, slightly affect the planet-planet encounter
properties.

For the same reason, the total number of FFPs does not affect the evolution of the stellar population in the star cluster. It simply scales the total number of close encounters and capture events involving planets. Inspired by the findings of \cite{sumi2011} we adopt an initial planet-to-star ratio of $\ratio(0)\equiv\np(0)/\nsb(0)=2$ in our default model, and also carry out simulations with $\ratio(0)=0.5$, and $\ratio(0)=1$, for comparison.

As planets are presumed to have been ejected from the planetary system in which they formed, the velocity dispersion of the FFP population is larger than that of the stellar population. The initial velocity of each FFP therefore constructed by combining two velocity vectors. The main component, $\vec{v}_p(r)$, representing the velocity of the host star at the moment of ejection, is drawn from the the Plummer distribution (see Eq.~\ref{eq:velocitydispersion}). The second velocity component, $\vec{v}_e$, is drawn from the distribution of ejection velocities at which the FFP escaped from its host star:
\begin{equation} \label{eq:totalvffp}
  \vec{v}_{\rm FFP}(r) = \vec{v}_p + \vec{v}_e \ .
\end{equation}
The resulting initial velocity dispersion for the FFP population is then
\begin{equation} \label{eq:supervirial}
  \sigma_{\rm FFP}^2 = \sigma_p^2(r) + \sigma_e^2 \ ,
\end{equation}
because we draw $\vec{v}_p$ and $\vec{v}_e$
independently and with random spatial orientations.

The (one-dimensional) velocity dispersion of the stars at the half-mass radius (Eq.~\ref{eq:velocitydispersion}) ranges between $\sigma_p(\rhm)\approx 0.3$~\kms{} (for $\nsb=500$ and $\rhm=1.54$~pc) and $\sigma_p(\rhm) \approx 1.5$~\kms{} (for $\nsb=3000$ and $\rhm=0.38$~pc). The planet ejection velocities are typically in the range $0.1-5$~\kms, and can therefore contribute substantially to the velocity dispersion of the FFPs in the star clusters. We therefore expect a relatively large fraction of the FFPs to escape at early times from low-density star clusters.


\subsection{Simulation parameters}

We use the NBODY6 package \citep{aarseth1999,aarseth2003} to carry out the simulations. Stellar evolution is modelled following the prescriptions of \cite{eggleton1989,eggleton1990} and \cite{hurley2000}.
The NBODY6 code was modified to suit
 the requirements of (i) the inclusion of the free-floating planets, and (ii)
 identification of the close encounters. 

In a star cluster with a virial radius $\rvir$ consisting of $N$ single stars, when a pair of stars approaches within a distance $r_{\rm close}\approx 4\rvir/N$ their velocities are significantly perturbed \cite[e.g.,][]{aarseth2003}. In our grid of simulations, $r_{\rm close}$ is in most cases several hundred to somewhat more than a thousand astronomical units. Inspired by the previous work of \cite{malmberg2007_encounters}, we record encounters within a distance of $r_{\rm enc} \leq 1000$~AU. Encounters with hard (i.e., tight) binaries are registered as such, while for encounters with soft (i.e., wide) binary systems, the encounters with individual stars are registered as encounters with single stars. The individual components of binary systems with periastron separations smaller than 1000~AU experience encounters on a regular basis, according to this definition, but are excluded from the analysis of the close encounter properties. All simulations are carried out until the star cluster has completely dissolved,
such that we are able to register all close encounters that occur during the
lifetime of each star cluster. In order to account for statistical effects, we simulate ten realisations for each model.


\section{Results} \label{section:results}


\subsection{Star cluster evolution and membership}

\subsubsection{Star cluster dynamics}

\begin{figure}
  \centering
  \includegraphics[width=0.45\textwidth,height=!]{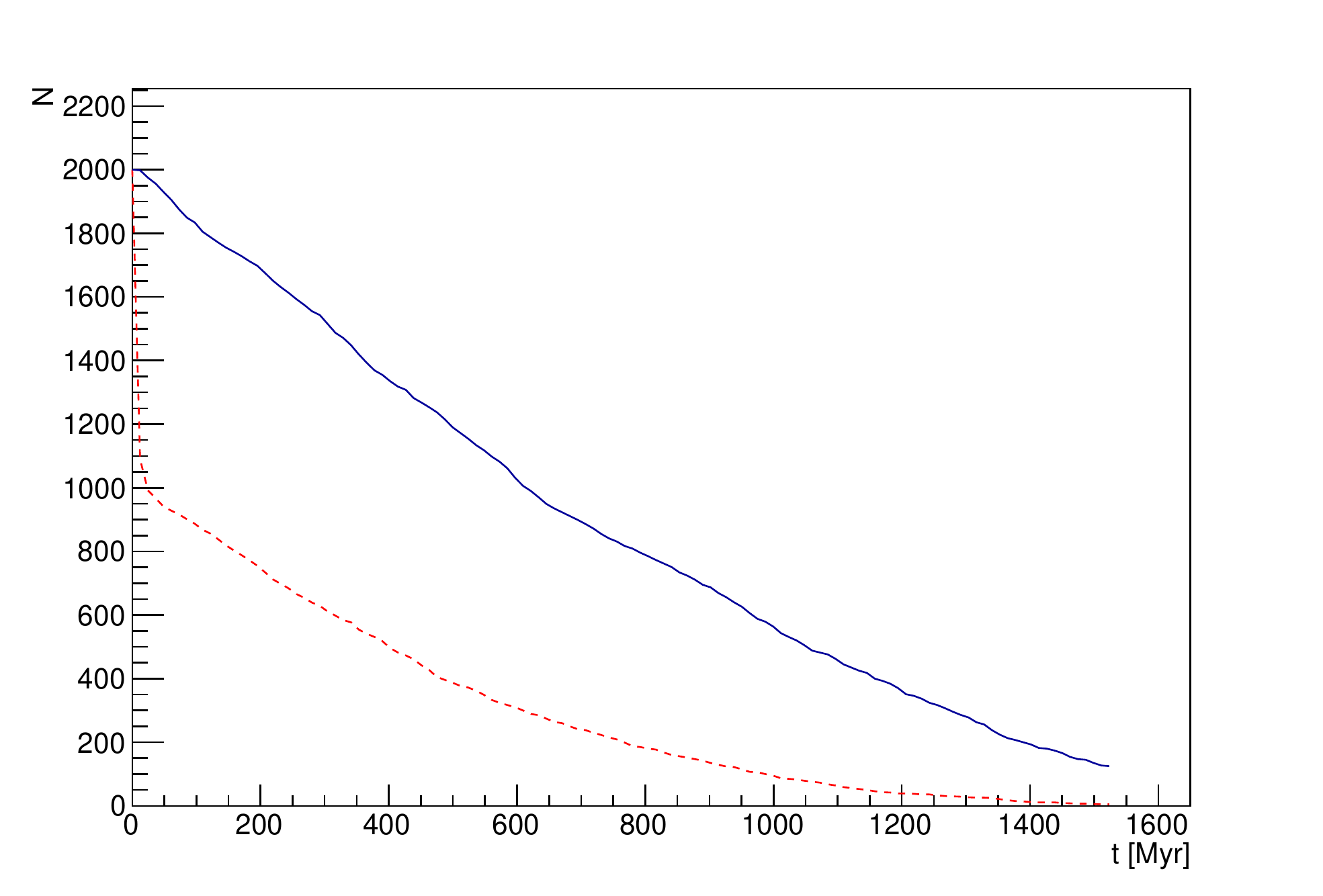}\\
  \includegraphics[width=0.45\textwidth,height=!]{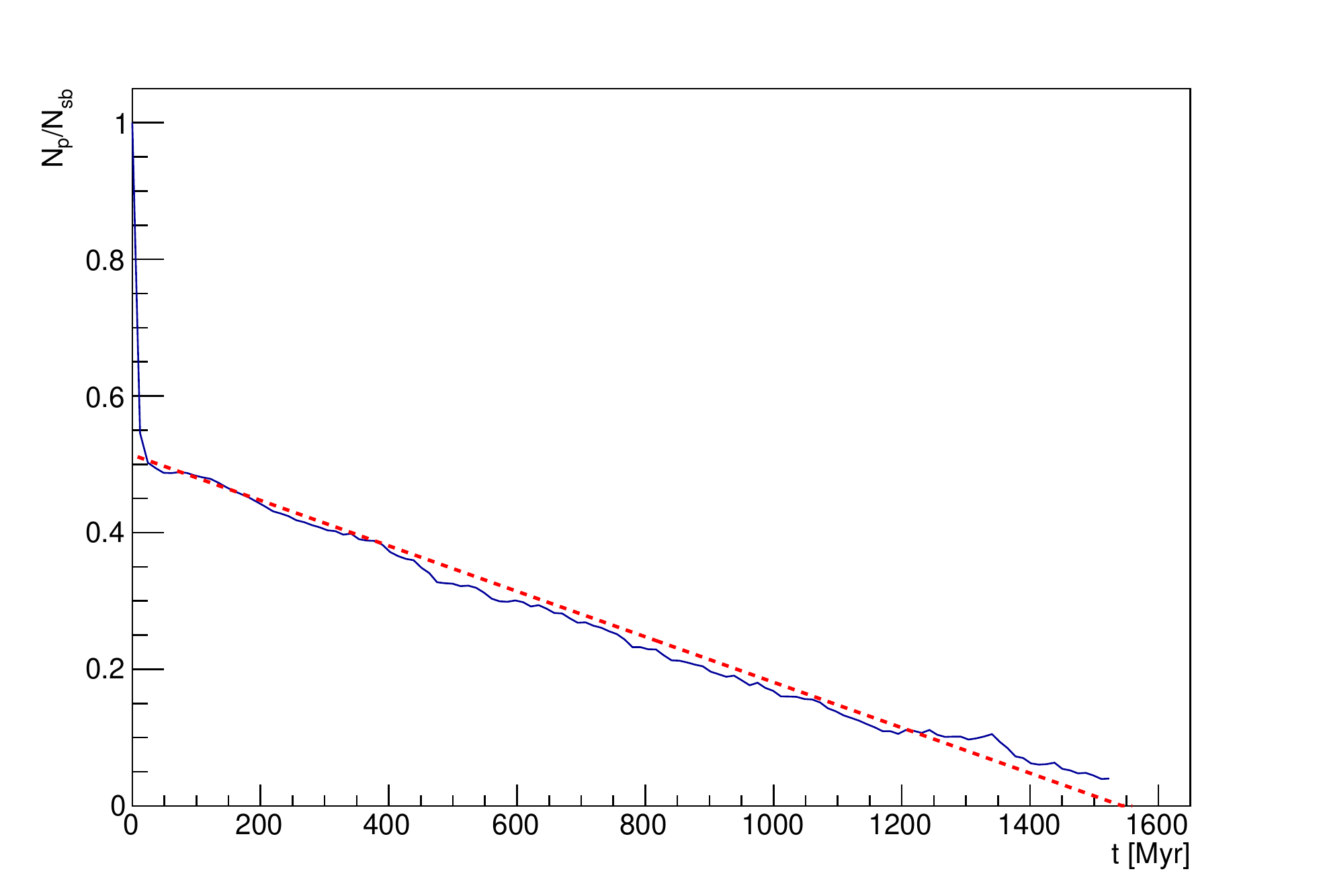}\\
  \caption{{\em Top}: the evolution of the number of cluster member stars $\nsb$ (blue
    curve) and planets $\np$ (red curve) as a function of time. This particular example shows the results for the model with $\nsb=2000$, $\binfrac=0\%$, $\ratio(0)=1$, and $\rhm=0.38$~pc. The distribution at which the FFPs are assumed to have been ejected from their host star is identical to the model with $\velocitytype=P$ (for $1\msun$) in Figure~\ref{figure:ejectionvelocities}.
        {\em Bottom}:
    the planet-to-star ratio $\ratio(t)=\np(t)/\nsb(t)$ as a function of time $t$. The red line indicates the best linear fit (Eq.~\ref{eq:ratiodefinition}).
  \label{figure:members} }
\end{figure}

\begin{figure}
  \centering
  \includegraphics[width=0.53\textwidth,height=!]{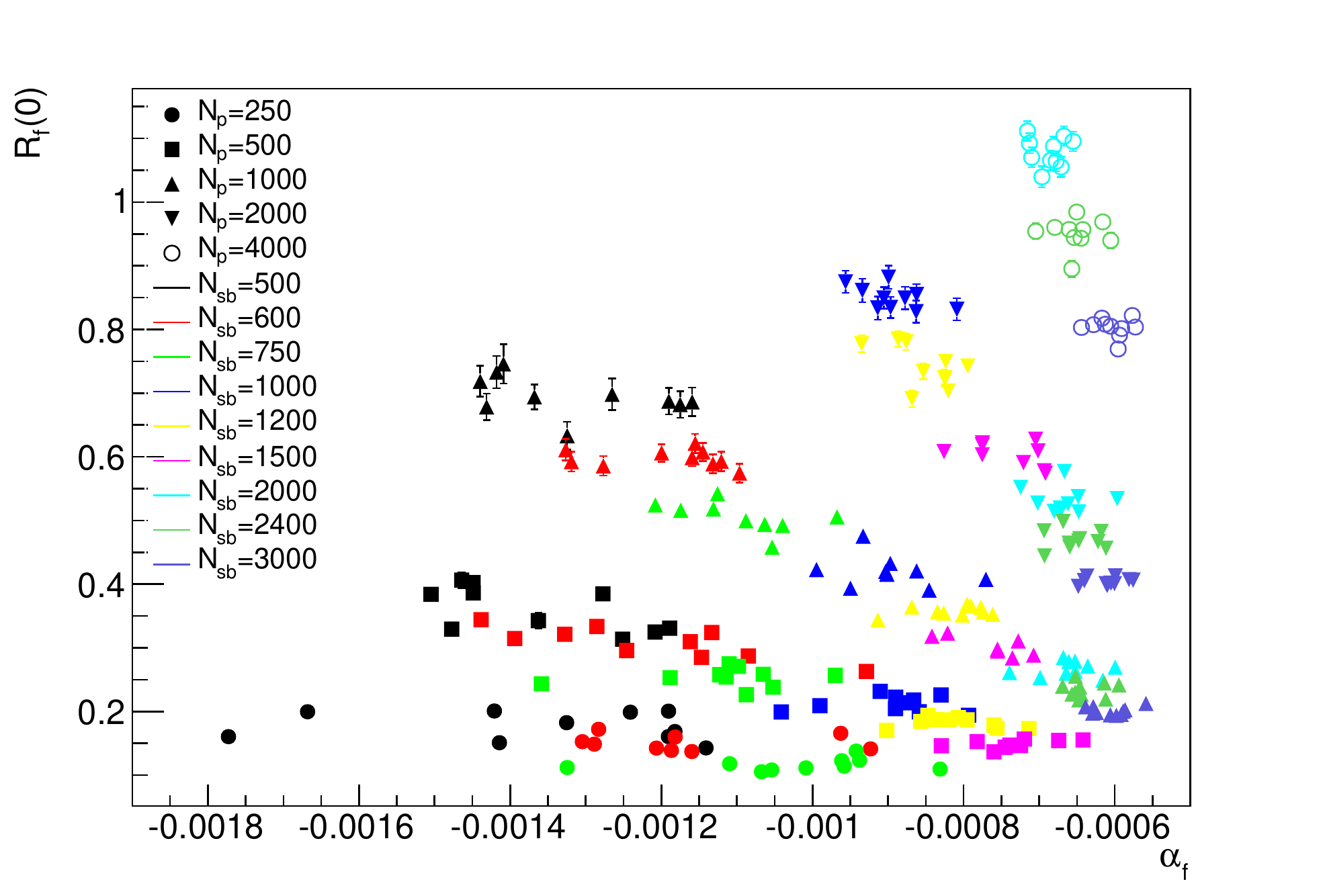}\\[-47ex]
  \quad\quad\quad\quad\quad(a) $\rhm=0.38~{\rm pc}$\\
  \quad\quad\quad\quad\quad\quad\quad$\velocitytype=P$ ($1\msun$)\\[40ex]
  \includegraphics[width=0.53\textwidth,height=!]{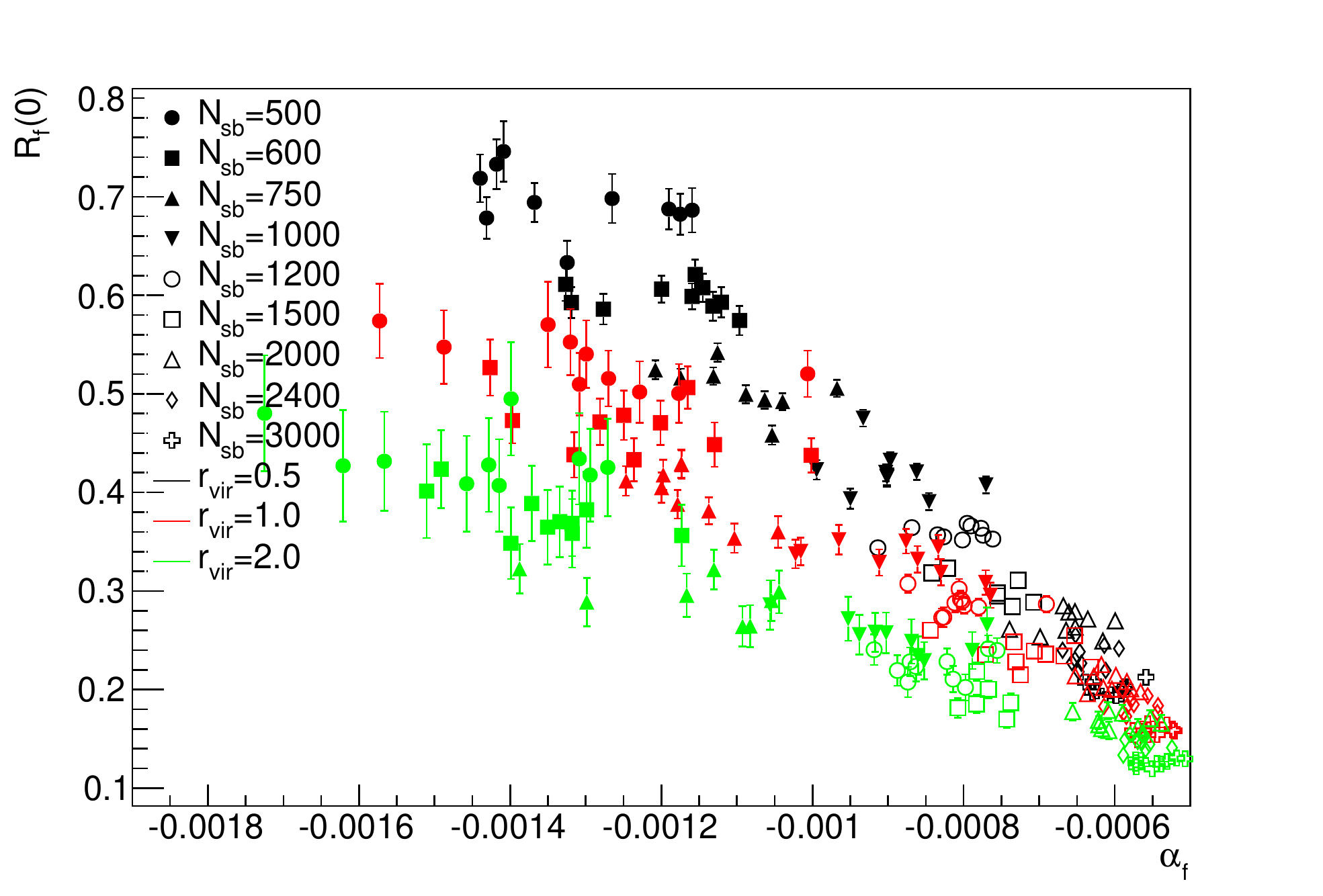}\\[-47ex]
  \quad\quad\quad\quad\quad\quad\quad\quad\quad\quad\quad\quad\quad\quad\quad(b) $\np=1000$\\
  \quad\quad\quad\quad\quad\quad\quad\quad\quad\quad\quad\quad\quad\quad\quad\quad\quad\quad\quad $\velocitytype=P$ ($1\msun$)\\[40ex]
  \includegraphics[width=0.53\textwidth,height=!]{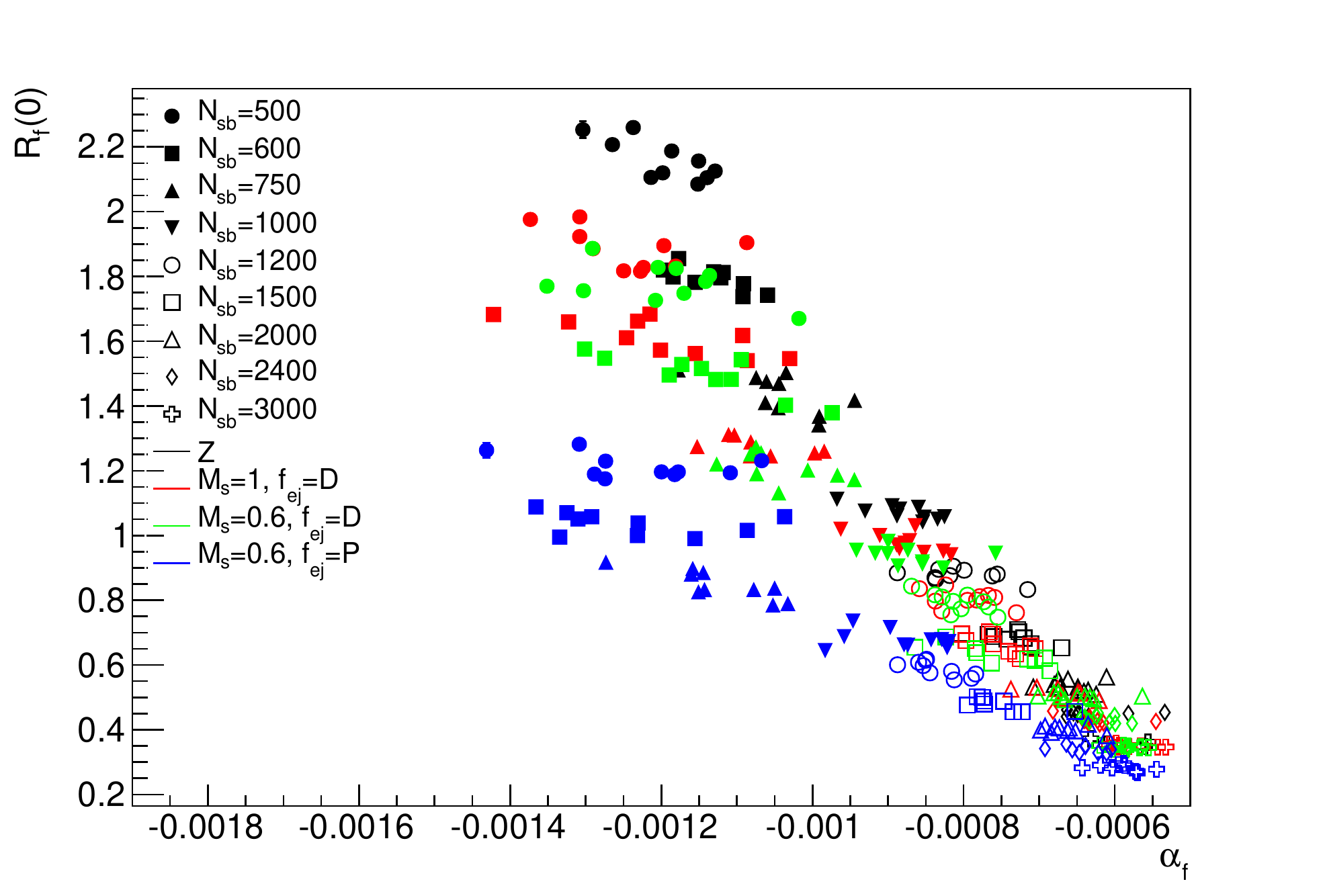}\\[-47ex]
  \quad\quad\quad\quad\quad\quad\quad\quad\quad\quad\quad\quad\quad\quad\quad\quad\quad(c) $\rhm=0.38~{\rm pc}$\\
  \quad\quad\quad\quad\quad\quad\quad\quad\quad\quad\quad\quad\quad\quad\quad\quad\quad$ \np=1000$\\[40ex]
  \caption{The fitted linear parameters $\ratiofit(0)$ vs. $\alphafit$ for planet-to-star ratio as a function of time (Eq.~\ref{eq:ratiodefinition}) for all models. {\em Top}: fixed $\rhm$ and ejection velocity distribution of planets, for clusters with different $\nsb=\ns+\nb$ and $\np$. {\em Middle}: fixed $\np$ and ejection velocity distribution of planets, for different $\rhm$ and $\nsb$. {\em Bottom}: fixed $\rhm$ and $\np$, for different planet ejection velocity distributions and different $\nsb$. The adopted initial planet ejection velocity distributions include prompt ejection ($P$), delayed ejection ($D$), and zero ejection velocity ($Z$); see Figure~\ref{figure:ejectionvelocities}. The symbols and their colours represent the different initial conditions for the simulations, as indicated in the legends. Each symbol represents the average result for an ensemble of star clusters.
\label{figure:memfit} }
\end{figure}

\begin{figure}
  \centering
    \includegraphics[width=0.53\textwidth]{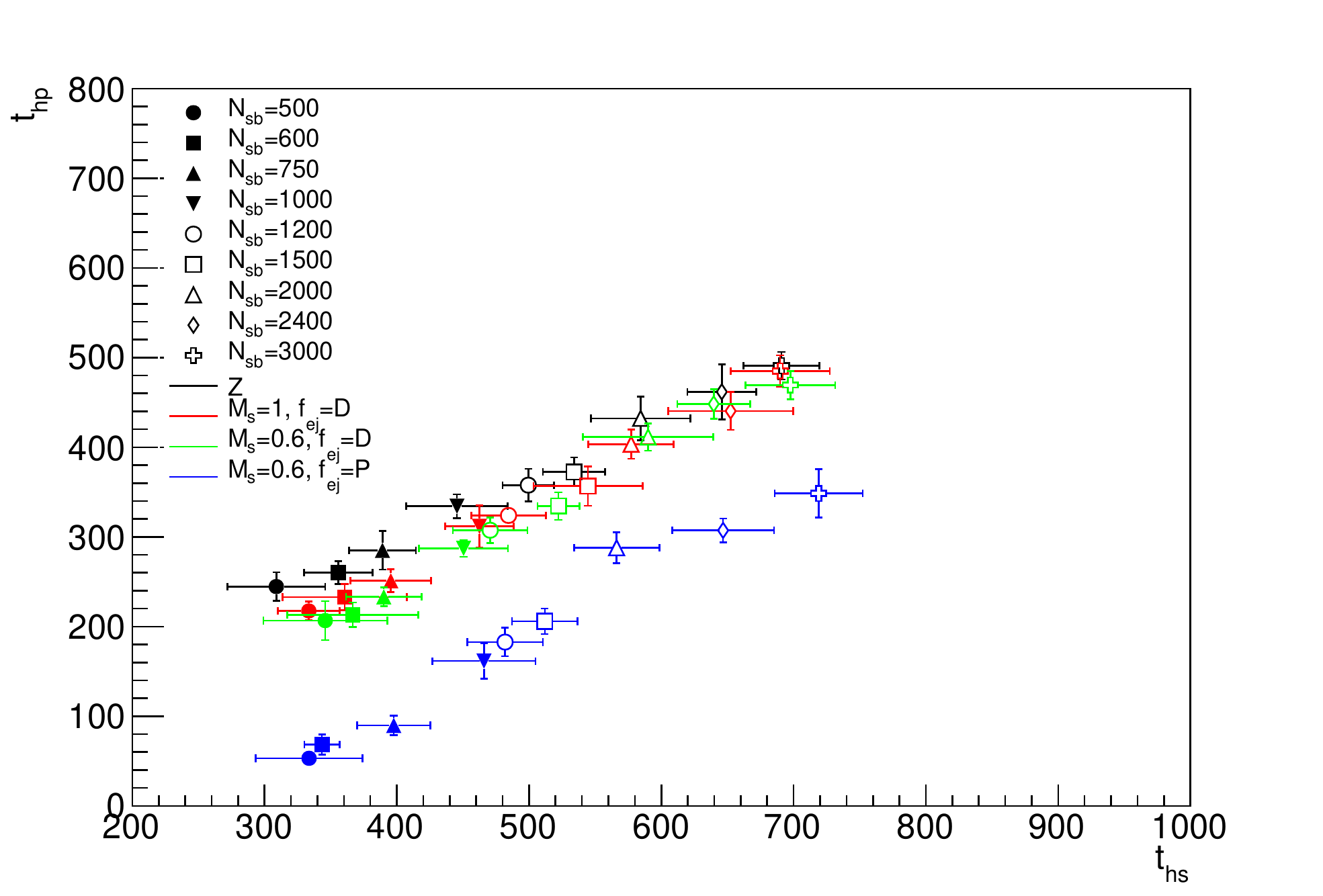}\\[-47ex]
    \quad\quad(a) $\rhm=0.38$~pc \\ [43ex]
    \includegraphics[width=0.53\textwidth]{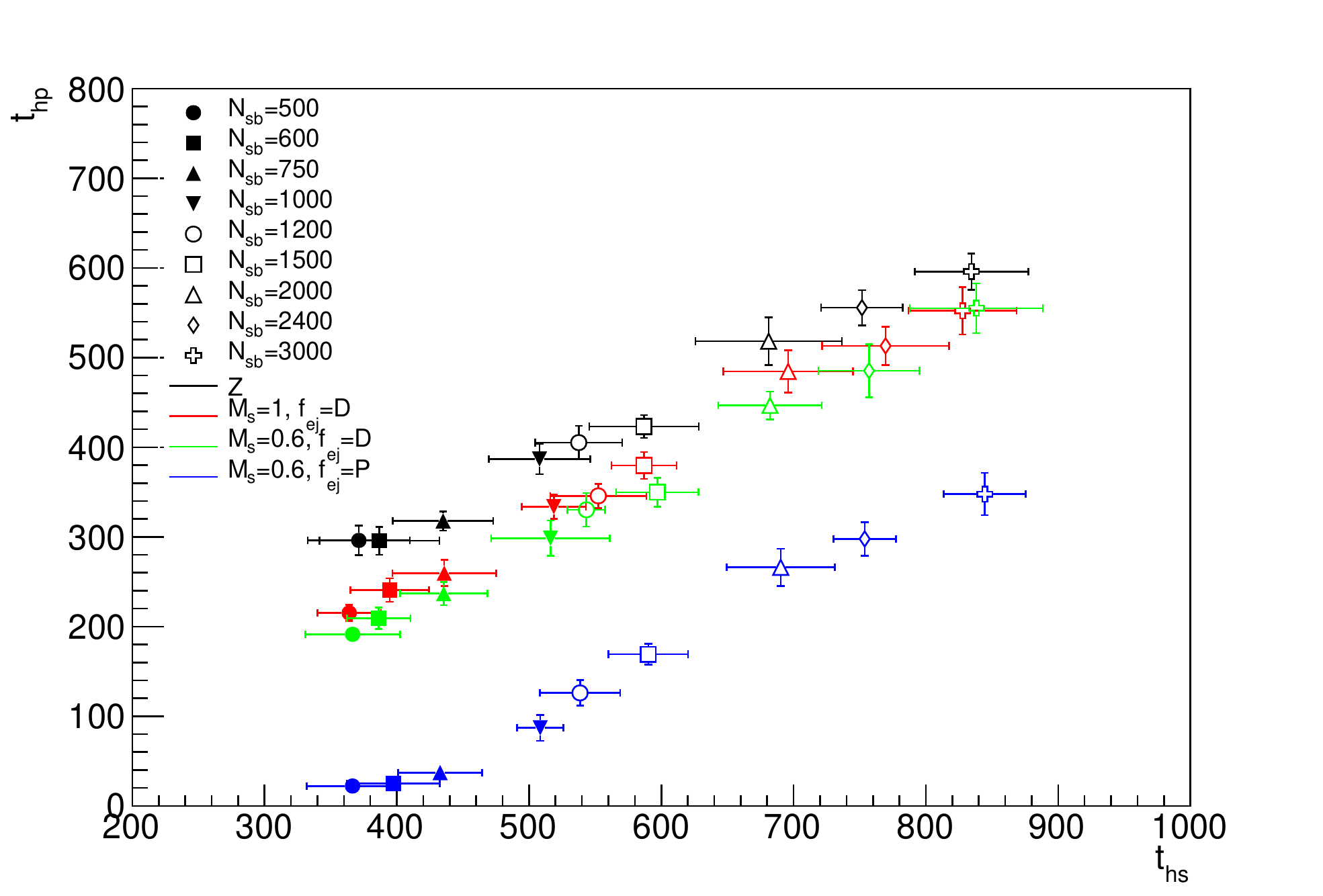}\\[-47ex]
    \quad\quad(b) $\rhm=0.77$~pc \\ [43ex]
    \includegraphics[width=0.53\textwidth]{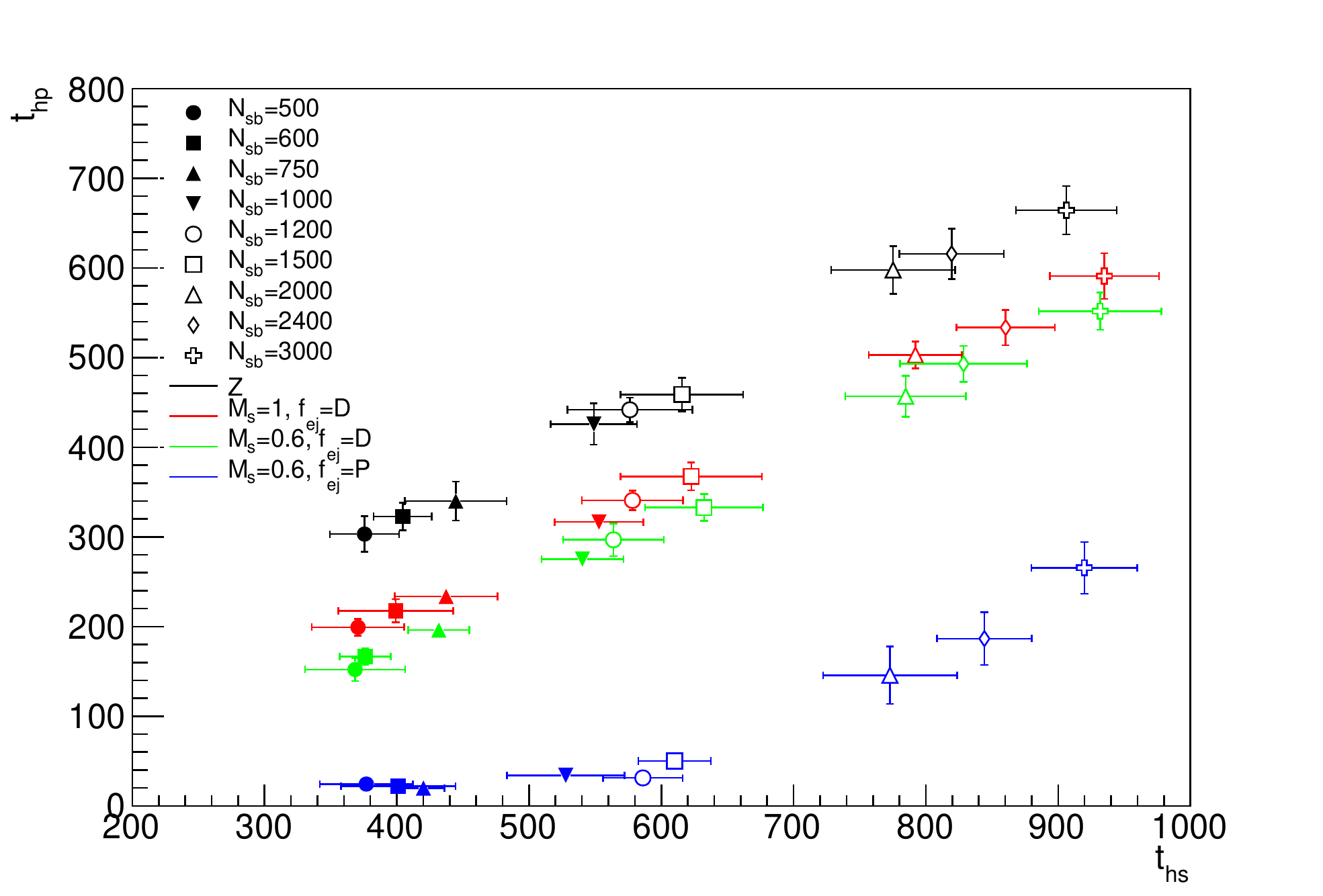}\\[-47ex]
    \quad\quad(c) $\rhm=1.54$~pc \\ [43ex]
  \caption{The relation between the half-number time $\thalfplanet$ of the number of planets in the clusters and the half-number time $\thalfstar$ of the number of stars in the clusters, for different sets of star clusters models and ejection velocity distributions. All models have $\np=1000$. The three panels show the results for half-mass radii of $\rhm=0.38$, 0.77, and 1.54~pc. Each symbol represents the average result for an ensemble of ten star clusters with a certain $\nsb=\ns+\nb$. The colours indicate the different initial planet ejection velocity distributions for prompt ejection ($P$), delayed ejection ($D$), and zero ejection velocity ($Z$); see also Figure~\ref{figure:ejectionvelocities}.\label{figure:thf} }
\end{figure}


The dynamical evolution of the stellar population in the star cluster is practically unaffected by the presence of the planets, as the planets are three
orders of magnitude lower in mass than the stars, and therefore effectively behave as test particles. The evolution of the stellar and planetary populations over time is illustrated in Figure~{\ref{figure:members}}, which shows that the number of FFP members in the cluster decreases faster than the number of stellar members. The reason for this twofold: (i) the FFP population is initially to some degree supervirial (see Eq.~\ref{eq:supervirial}), which results in the rapid escape of many of the FFPs with a high initial velocity, and (ii) dynamical interactions result in FFPs obtaining higher velocities than stellar components, resulting in the preferential loss of planets.

The initial escape velocity $v_{\rm esc}$ for the Plummer model as a function of distance to the cluster centre $r$ is
\begin{equation} \label{eq:escapevelocity}
  v_{\rm esc}(r) = \left( \frac{2GM}{r} \right)^{1/2}  \left( 1+\frac{\rhm^2}{r^2} \right)^{-3/4}  
\end{equation}
\citep{heggiehut}, where $M$ is the total cluster mass and $G$ the gravitational constant. At the half-mass radius ($r=\rhm$) this equation reduces to $v_{\rm esc} \approx 0.84 \sqrt{GM/\rhm}$. For our models its value typically ranges between $v_{\rm esc}(\rhm)\approx 0.7$~\kms{} ($\nsb=500$ and $\rhm=1.54$~pc) and $v_{\rm esc}(\rhm) \approx 3.6$~\kms{} ($\nsb=3000$ and $\rhm=0.38$~pc). Although a number of FFPs are able to escape the cluster immediately, the FFPs with a relatively small ratio $\sigma_e/\sigma_p(r)$ are able to remain part
of the star cluster for some time, but are typically removed at earlier times than the stars. The stellar population is barely affected by these encounters, but the typical velocity of the FFPs in the star cluster rapidly increases. Two-body encounters  typically occur on a relaxation time. For stars with a mass close to the average stellar mass in a Plummer sphere, the (initial) relaxation time $\trelax$ at the half-mass radius is 
\begin{equation}  \label{eq:relaxationtime}
  \trelax = \frac{0.206\nsb\rhm^{3/2}}{\sqrt{GM}\ln \Lambda} \ ,
\end{equation}
where $\nsb$ is the number of stellar-mass objects, $M=\nsb\langle\mstar\rangle(1+\binfrac)$ the total cluster mass, and $\ln\Lambda \approx \ln \nsb$ is the Coulomb logarithm \citep{binneytremaine, heggiehut}. The half-mass relaxation time in our simulations ranges from  $\trelax \approx 7$~Myr (for $\nsb=3000$ and $\rhm=0.38$~pc) to $\trelax\approx 30$~Myr (for $\nsb=500$ and $\rhm=1.54$~pc). 

Encounters between the star cluster members also result in energy equipartition, also roughly on a timescale 
\begin{equation}
	t_{\rm ms} = \left( \frac{\langle \mparticle \rangle}{M_{\rm max}} \right)\trelax \ ,
\end{equation}
\citep[e.g.,][]{spitzer1987} where $\langle \mparticle \rangle=\langle\mstar\rangle(1+\binfrac)$ is the average mass of a particle participating in the mass segregation process (a single star or a binary star) and $M_{\rm max}$ is the mass of the most massive particle in the cluster. In our simulations, $\langle \mstar \rangle / M_{\rm max} \approx 0.1$, and the mass segregation timescale therefore ranges from 0.3~Myr to 3~Myr for the set of models studied in this paper \citep[but may be shorter if the star clusters are initially substructured, see][]{allison2009a, allison2009b}. These timescales, as well as the dissolution time of the star clusters, also depend on the range of stellar masses present in the cluster \citep[e.g.,][]{kouwenhoven2014}. FFPs do not contribute to the mass segregation process in the sense that they do not affect the stellar population. However, they do obtain higher velocities as a result of close encounters with stars, also roughly on a timescale $\trelax$.


\subsubsection{Analytic estimates for the survival times}

The FFP population can be separated into three categories: (i) FFPs that escape the cluster immediately, (ii) FFPs that are ejected from the cluster as a result of strong three-body encounters, and (iii) FFPs that evaporate from the cluster following a series of weak encounters and an interaction with the Galactic tidal field. Whether or not a FFP can escape from the star cluster immediately depends on a combination of factors, but most importantly the ratio between the FFP's initial ejection velocity and the escape velocity at the initial location of the FFP. Since we set up the FFP planets following a Plummer distribution, the initial cumulative number of planets $N(r)$ as a function of distance $r$ to the cluster centre is given by:
\begin{equation} \label{eq:planetdistribution}
  N(r) = N \left( 1+\frac{\rhm^2}{r^2} \right)^{-3/2} \ ,
\end{equation}
where $N$ is the total number of planets in the system. The number of FFPs that escape at early times can then be obtained by combining Eqs.~\ref{eq:escapevelocity} and~\ref{eq:planetdistribution}, and integrating over the entire cluster. Escapers are removed  when they reach a distance of $2r_t$ (Eq.~\ref{eq:tidalfield}). For high-velocity FFPs this escape time is roughly $2r_t(M)/\sigma_{\rm FFP}\approx 1-10$~Myr.

It is  possible to estimate the evolution of the FFPs beyond the initial phase of escape analytically. The linear relation between $\thalfstar$ and $\thalfplanet$ in Figure~\ref{figure:thf} is remarkable, and can be explained as follows. The number of stellar members of the cluster decreases with time, and this trend can to first order be approximated as linear:
\begin{equation} \label{eq:starloss}
	\nsb(t)= \nsb(0)\left( 1-\frac{t}{\tdiss}\right) 
\end{equation}
\citep[e.g.,][]{baumgardt2003, heggiehut, lamers2005}.
With this simple relation, the time at which half of the stars have escaped from the star cluster is $\thalfstar=\tdiss/2$. The planet-to star ratio $\ratio(t)\equiv \np(t)/\nsb(t)$ decreases rapidly in a certain interval $0<t<t_1$ as high-velocity planets escape. Beyond this time, a certain fraction $x\equiv \np(t_1)/\np(0)$ of the FFPs remains. The expression for $\ratio(t)$ for $t>t_1$ is thus
\begin{equation} \label{eq:ratiostarloss}
	\ratio(t) = \frac{\np(t_1)}{\nsb(t_1)}  \left(\frac{\tdiss-t}{\tdiss-t_1}\right) \ .
\end{equation}
The number of FFPs in the star cluster as a function of time can then be computed using Eqs.~\ref{eq:starloss} and~\ref{eq:ratiostarloss}:
\begin{equation} \label{eq:solution}
	\begin{array}{ll}
	\np(t) & = \ratio(t)\nsb(t) \\
	       & =\frac{\np(t_1)\nsb(0)}{\nsb(t_1)}\left(\frac{\tdiss-t}{\tdiss-t_1}\right) \left(1-\frac{t}{\tdiss}\right) \\
	       & = \np(t_1)\left(\frac{\tdiss-t}{\tdiss-t_1}\right)^2 \ .
	\end{array}
\end{equation}
The time $\thalfplanet$ at which half of the planets have escaped from the star cluster can then be obtained by solving Eq.~\ref{eq:solution} for $\np(t)/\np(0) = 1/2$:
\begin{equation} \label{eq:halfplanethalfstar}
	\thalfplanet = \thalfstar\left( 2 - \sqrt{\frac{2}{x}} \right) + t_1\sqrt{\frac{1}{2x}} \ ,
\end{equation}
where $x\equiv\np(t_1)/\np(0)$ indicates the fraction of FFPs that remain bound to the star cluster during the time interval $[0,t_1]$. The above equation is valid for $\np(t_1)/\np(0)>1/2$, as larger initial escape fractions result in $\thalfplanet<t_1$. In the limit where all FFPs are initially virialized, so that $x=1$ and $t_1=0$, Eq.~\ref{eq:halfplanethalfstar} reduces to
\begin{equation}
	\thalfplanet = (2-\sqrt{2})\thalfstar \approx 0.59\thalfstar
\end{equation}
In other words, if the FFP population in a star cluster initially has a velocity dispersion similar to the velocity dispersion of the star cluster, the planets typically escape at 60\% of the remaining lifetime of the star cluster. This result is independent of the star cluster parameters, and provides a good estimate of the fate of a planetary population, under the condition that Eq.~\ref{eq:starloss} is a reasonable approximation. 

From Figure~\ref{figure:thf} we can estimate the fraction of planets $(1-x)$ that escapes during the initial time $0<t<t_1$ by solving Eq.~\ref{eq:halfplanethalfstar} for $x$:
\begin{equation} \label{eq:x}
	x = \frac{1}{2}\left(\frac{2\thalfstar-t_1}{2\thalfstar-\thalfplanet}\right)^2 \approx \frac{1}{2}\left(\frac{2\thalfstar}{2\thalfstar-\thalfplanet}\right)^2 \ .
\end{equation}
This estimate is only valid when all FFPs are generated at $t=0$, which is generally not the case in realistic star clusters. However Eqs.~\ref{eq:halfplanethalfstar} and~\ref{eq:x} but it can also be used to statistically estimate long a FFP typically remains a member of a star cluster.

The results in Figure~\ref{figure:thf} are well described by Eq.~\ref{eq:halfplanethalfstar}. The relation between $\thalfplanet$ and $\thalfstar$ is indeed linear, although slightly steeper in the regime where the star clusters can retain the majority of the FFPs for a substantial amount of time. The relation breaks down when many FFPs escape at early times. The time $\thalfplanet$ depends strongly on the ejection velocity distribution of the FFPs. A comparison between the panels in Figure~\ref{figure:thf} also shows that the FFPs escape earlier from clusters with a smaller velocity dispersion (i.e., a larger $\rhm$). For $\velocitytype=P$ ($1\msun$) the ejection velocity of the planets is largest, and therefore $\thalfplanet$ is small; many planets escape within $10-20$~Myr.


\subsubsection{Escape of the planet population}

Due to the preferred ejection of FFPs from the system, the planet-to-star ratio $\ratio(t)\equiv \np(t)/\nsb(t)$ rapidly decreases over time. The evolution of $\ratio(t)$ is roughly linear for most of the time, and we therefore quantify this evolution with a linear fit:
\begin{equation} \label{eq:ratiodefinition}
  \ratio(t) \equiv \frac{\np(t)}{\nsb(t)} \approx \ratiofit(0) \left[ 1 + \alphafit t \right] \ ,
\end{equation}
where $t$ is the time in units of Myr and $\alpha$ in units of Myr$^{-1}$. In the ideal case where this dependence is truly linear, the quantity $\ratiofit(0)$ is the initial planet-to-star-ratio, and the quantity $\alphafit \approx d\ratio(t)/dt$ indicates the relative escape rate of planets in units of Myr$^{-1}$. Although this derivative varies slightly with time, it is to first order constant. Under this assumption, the timescale at which all planets are removed from the star cluster is $\tdiss \approx -\alphafit^{-1}$. Apart from statistical fluctuations, $\alphafit$ is independent of the initial number of FFPs, $\np$. Since the number of planets decreases more quickly than the number of stars, $\alphafit$ is always negative. The quantity $\alphafit$ can be used to estimate the number of planets in the star cluster at any time. 

In the case of a FFP population that is initially close to virial equilibrium with the stellar population, which is the case in Figure~\ref{figure:members}, the fitted value $\ratiofit(0)$ is close to (but not equal to) the initial planet-to-star ratio. When planets are ejected from their host systems with high velocities, a large number escape immediately, and those in the low-velocity tail of the distribution follow Eq.~\ref{eq:ratiodefinition}. Due to early escape, the fitted value $\ratiofit(0)$ is smaller than the initial planet-to-star ratio $\ratio(0)$. This difference is thus an indicator of the initial fraction of high-velocity FFPs.

The fitted parameters $\ratiofit(0)$ and $\alphafit$ for the different models are plotted in Figure~{\ref{figure:memfit}}. The top panel shows the values for the models with $\rhm=0.38$~pc and $\velocitytype=P$ ($1\msun$). All models have $-1.4\times 10^{3} < \alphafit < -0.5\times 10^{-3}$, which indicates that all planets are ejected from the clusters on timescales of roughly $0.7-2$~Gyr. For models with identical values of $N$ the scatter in $\ratiofit(0)$ is small and concentrates around the initial planet-to-star ratio $\ratio(0)$. The time scale $\tdiss$ at which all planets escape from the cluster increases with increasing $\nsb$, which corresponds to the dependence of the relaxation time $\trelax$ on $\nsb$ (Eq.~\ref{eq:relaxationtime}). In addition, the results are (apart from statistical differences) independent of the number of FFPs, $\np$.

Figure~{\ref{figure:memfit}b} shows a comparison for clusters with identical $\np$ and $\velocitytype$, but with a varying number of stars $\nsb$ (different symbols) and half-mass radii $\rhm$ (different colors). Initially, $\ratio(0)=\np/\nsb$ varies between 0.33 and 2. The ejection timescale for the FFPs ranges from $\tdiss \approx 700$~Myr for $\nsb=500$ to $\tdiss=2$~Gyr for $\nsb=3000$. Larger values of $\rvir$ result in a smaller fitted value  $\ratiofit(0)$ as compared to the initial conditions, as more FFP speeds are initially above the escape velocity (Eq.~\ref{eq:escapevelocity}) and therefore escape at early times.
For the clusters with the largest number of particles (where most FFPs are initially below the escape velocity), $\alphafit$ increases with increasing $\nsb$, indicating that clusters with a smaller radius eject their planets faster, due to their shorter relaxation time.

Figure~\ref{figure:memfit}c shows the dependence of $\ratiofit(0)$ for star clusters with identical $\rhm$ and $\np$, but with different numbers of stars $\nsb=N(1+\binfrac)$ and ejection velocity distributions $\velocitytype$. After this initial phase of escape of high-velocity FFPs, the ratio $\ratio(t)$ decreases roughly linearly with time. As this initial phase is short, and as $\alpha$ is independent of $\np$ beyond this time, the corresponding timescale for the escape of the entire planetary population is still well-approximated with $\tdiss=-\alphafit^{-1}$, irrespective of the choice for $\velocitytype$.

The time at which the planet-to-star ratio $\ratio(t)$ drops to half the value it had {\rm after the initial rapid escape phase}, is $\tdiss/2$. 
We quantify this half-life time for the stellar population with $\thalfstar$, and for the FFP population with $\thalfplanet$. 
The relation between $\thalfstar$ and $\thalfplanet$ for the different star clusters is shown in Figure~\ref{figure:thf}. From top to bottom, the three panels show the results for models with $\rhm=0.38$~pc, 0.77~pc, and 1.54~pc, respectively. The initial number of planets in each cluster is $\np=1000$. Since planets escape faster due to their initially supervirial state and/or two-body relaxation, all star clusters have $\thalfstar > \thalfplanet$, and both quantities are independent of $\np$. Note that the choice for $\np$ does not affect the results in the figure, as the planets effectively behave as test particles.


\subsection{Close encounters}

During the simulations we record all close ($< 1000$~AU) encounters between the members of the star cluster. Two-body encounters can be of the type star-star (S-S), star-planet (S-P), and planet-planet (P-P). The effect of gravitational focusing for S-S encounters is largest, while it is almost negligible for P-P. Three-body encounters occur regularly and almost exclusively involve stellar binary systems (SS-S and SS-P). Three-body encounters that result in the formation of a dynamical binary system almost always result in the ejection of the body with the lowest mass. The formation of an SS binary is therefore much more common than a binary of type SP (a planet orbiting a star), while those of type PP (binary planets) only rarely occur. Moreover, FFPs tend to escape earlier than stars. As a result, close encounters of type SP-S are more common than those of type SP-P, even when initially $\ns<\np$. Three-body encounters involving more than one planet in a bound system (SP-P and PP-S) are rare, since this first requires the formation of these systems by dynamical capture. As initially all planets in the star clusters are free-floating\footnote{Close encounters involving star-planet system (SP) may be common in realistic star clusters as a large fraction of stars is likely to host one or more planets. In our simulations, however, all stars are initialised without planetary companions.}, the encounter frequencies for the interactions SS-SP, SS-PP, SP-PP and PP-PP are negligible, primarily because capture of FFPs into SP-type systems is rare, while double planets (PP) almost never form. The only four-body encounters that occur frequently are those between two stellar binary systems (SS-SS), particularly when primordial binaries are present. Despite these arguments, it should be noted that planetary systems may be common in real star clusters, and our results are therefore lower limits for the encounter frequencies of planet-hosting stars in open clusters.

The encounter frequencies at a given time can be estimated using the abundances of the different types of systems. Under the assumption that all particles have a similar distribution in phase-space, and ignoring the effects of gravitational focusing, the number of encounters $\nenc$ of the different types can be estimated through the following proportionalities:
\begin{equation} \label{eq:encounterrates}
	\begin{array}{lll}
	\nenc_{S-S}   & \propto \ns^2   & = (1-\binfrac)^2N^2\\
	\nenc_{S-P}   & \propto \ns\np  & = (1-\binfrac)N\np \\
	\nenc_{P-P}   & \propto \np^2 \\
	\nenc_{SP-S}  & \propto \nsb\ns\np & = (1-\binfrac^2)N^2 \np \\
	\nenc_{SS-S}  & \propto \ns\nb  & = (1-\binfrac)\binfrac N^2 \\
    \nenc_{SS-P}  & \propto \nb\np  & = \binfrac N \np \\
    \nenc_{SS-SS} & \propto \nb^2   & = \binfrac^2 N^2 \quad\quad .\\
	\end{array}
\end{equation}
In addition, all encounter rates depend on the choice of the encounter criterion $\delta r < p_m$ and the size of the star cluster $\rvir$ through the approximate proportionality $\nenc \propto (p_m/\rvir(t))^3$ at time $t$.
These estimates are valid for short periods of time where the populations do not change substantially though the formation or destruction of gravitationally bound two-body systems or dynamical mass segregation.


\subsubsection{Encounter rates for dynamical populations} \label{section:encounterrates}

\begin{table*}
  \caption{The frequency of the various types of encounters during the lifetime of a subset of the star clusters. All values and their errors represent the average and the standard deviation for an ensemble of ten realisations. Several types of encounters are very rare (such as SS-SP), and are therefore not listed here. \label{table:encounterrates} }
    \begin{tabular}{p{0.0mm}|p{3mm}|p{0mm}|p{3mm}|p{2mm}|p{2mm}|p{2mm}|c|c|c|c|c|c|c|}
      \hline\hline
      ID & $N$&$\binfrac$&$\np$&$\rvir$&$M_{s}$&$\velocitytype$&S-S&S-P&P-P&SS-S&SP-S&SS-P&SS-SS\\
      &     &  \%   &     & pc    &$\msun$& & (1) & (2) & (3) & (4) & (5) & (6) & (10) \\
      \hline
1 & 500& 0&1000&0.5&1.0&P&$872\pm119$&$357\pm183$&$77\pm83$&$93\pm15$&$1\pm1$&$7\pm5$&$2\pm2$\\
2 &1000& 0&2000&0.5&1.0&P&$4374\pm461$&$2590\pm1139$&$656\pm648$&$254\pm31$&$2\pm2$&$17\pm8$&$7\pm3$\\
3 &2000& 0&4000&0.5&1.0&P&$22880\pm2105$&$20316\pm6745$&$6585\pm5374$&$707\pm87$&$7\pm5$&$57\pm17$&$13\pm4$\\
\hline
4 &2000&20&4000&0.5&1.0&P&$26265\pm2044$&$22959\pm7741$&$7700\pm5901$&$7599\pm796$&$29\pm13$&$2848\pm956$&$587\pm100$\\
5 &2000&50&4000&0.5&1.0&P&$33814\pm3462$&$28254\pm7824$&$9523\pm6539$&$18817\pm1555$&$90\pm23$&$7472\pm2144$&$2715\pm277$\\
\hline
6 &2000& 0&1000&0.5&1.0&P&$21931\pm1350$&$4723\pm1450$&$377\pm287$&$726\pm72$&$1\pm1$&$16\pm7$&$13\pm4$\\
7 &2000& 0&2000&0.5&1.0&P&$22458\pm2016$&$9796\pm3265$&$1602\pm1296$&$736\pm86$&$3\pm2$&$33\pm16$&$11\pm6$\\
\hline
8 &2000& 0&4000&1.0&1.0&P&$8190\pm745$&$4548\pm2073$&$1169\pm1232$&$457\pm56$&$1\pm1$&$28\pm9$&$14\pm6$\\
9 &2000& 0&4000&2.0&1.0&P&$3286\pm385$&$1051\pm593$&$190\pm241$&$255\pm34$&$1\pm1$&$9\pm9$&$7\pm3$\\
\hline
10&2000& 0&4000&0.5& -- &Z&$21916\pm1826$&$42707\pm2314$&$27029\pm1348$&$695\pm78$&$15\pm5$&$139\pm24$&$11\pm4$\\
11&2000& 0&4000&0.5&1.0&D&$21875\pm1670$&$36927\pm7276$&$20778\pm5758$&$694\pm77$&$10\pm7$&$119\pm29$&$13\pm4$\\
12&2000& 0&4000&0.5&0.6&D&$22070\pm2225$&$36351\pm3993$&$19765\pm3203$&$726\pm51$&$12\pm6$&$126\pm17$&$12\pm4$\\
13&2000& 0&4000&0.5&0.6&P&$19951\pm7343$&$25574\pm9194$&$10612\pm3806$&$627\pm227$&$8\pm5$&$98\pm36$&$9\pm5$\\
      \hline\hline
    \end{tabular}
\end{table*}

The frequency of occurrence of the various types of encounters for a subset of the modelled star clusters with different initial conditions are listed in Table~\ref{table:encounterrates}. The quantities represent averaged results for an ensemble of ten realisations of each model. Note these are cumulative values over the entire lifetime of the star clusters. The relations in Eq.~\ref{eq:encounterrates} therefore only provide first-order estimates. Model~3 is the reference model that is used for comparison with models that have different initial conditions.

Models $1-3$ represent the results for star clusters with different $N$ and identical initial planet-to-star ratios, and shows that the number of encounters $\nenc$ grows faster than the number of particles in the star cluster. This is because these three clusters have the same $\rhm$ and therefore a stellar density and a total lifetime that increase with $N$. For the high-density star clusters the ratio $\nenc_{P-P}/\nenc_{S-S}$ is substantially larger than for low-density star clusters. The reason for this is that the ejection velocity for the planets is identical for all models, and therefore planets in the low-mass star clusters are supervirial and escape at earlier times. Although the initial binary fraction is zero for these three models, several binary systems (SS and SP) form and have close encounters with the other members of the star clusters.

Models $3-5$ show the results for models with an identical $N=\ns+\nb$ and $\np$, but with different binary fractions $\binfrac$. The total number of individual stars in the clusters increases as $\nsb = \ns+2\nb = N(1+\binfrac)$, and the total cluster mass follows the same proportionality. The number of S-S and S-P encounters initially decreases with increasing $\binfrac$, but due to higher mass density, the destruction of binary systems and the longer lifetime of star clusters with a higher $\binfrac$, the cumulative number of these types of encounters increases with increasing $\binfrac$. For similar reasons, the P-P encounter rate, which is initially independent of $\binfrac$, increases with $\binfrac$ due to dependence of the dissolution timescale on $\binfrac$. The total number of SS-S, SS-P, and SS-SS encounters increase strongly with $\binfrac$, although the occasional formation and destruction of binary systems changes the proportionality of Eq.~\ref{eq:encounterrates} mildly.

The dependence of the the encounter frequencies on the number of planets $\np$ can by studied by comparing models~3, 6, and 7. As expected, the encounters involving only stars (S-S, SS-S, SS-SS) are independent of $\np$. The number of encounters between stars and planets (S-P, SS-P) are proportional to $\np$, while for the planet-planet encounters (P-P) the proportionality is $\np^2$. The number of SP-S encounters is well described by Eq.~\ref{eq:encounterrates} with proportionality constant $7.5\times 10^{-10}$.

Models 3, 8 and 9 represent identical star clusters apart from their initial size $\rvir$. Since the initial virial radius of model~8 is twice as large as as that of model~3, we expect all encounters involving only stars (S-S, SS-S, SS-SS) to occur roughly eight times less frequently. The number of encounters between FFPs occur even less frequently because the ratio between the planet ejection velocities and the stellar velocity dispersion becomes larger as $\rvir$ increases, resulting in the initial escape of a larger number of FFPs.

Models 3 and $10-13$ represent five star clusters with (statistically) identical stellar populations, but with different planet ejection velocity distributions $\velocitytype$. Model~10 represents the special case where all planets are assumed to be ejected from their host star with zero velocity, i.e., the FFP population in this model is initially in virial equilibrium. As expected, the number of encounters of types S-S, SS-S, SS-SS are statistically identical. All encounters involving FFPs depend on the ability of the star cluster to retain the escaping planetary population. The number of encounters involving FFPs (S-P, P-P, SS-P) increases when the clusters can retain the FFPs for longer times, such as when the typical ejection velocity (Figure~\ref{figure:ejectionvelocities}) is lower. The number of FFPs that are captured into SP-systems is largest when the ejection velocities are smallest. Models in which the FFPs have lower initial velocities therefore result in a larger number of encounters of type SP-S.


\subsubsection{The encounter number distribution} 

\begin{figure*}
  \centering
  \begin{tabular}{p{0.5\textwidth}p{0.5\textwidth}}
	\includegraphics[width=0.5\textwidth]{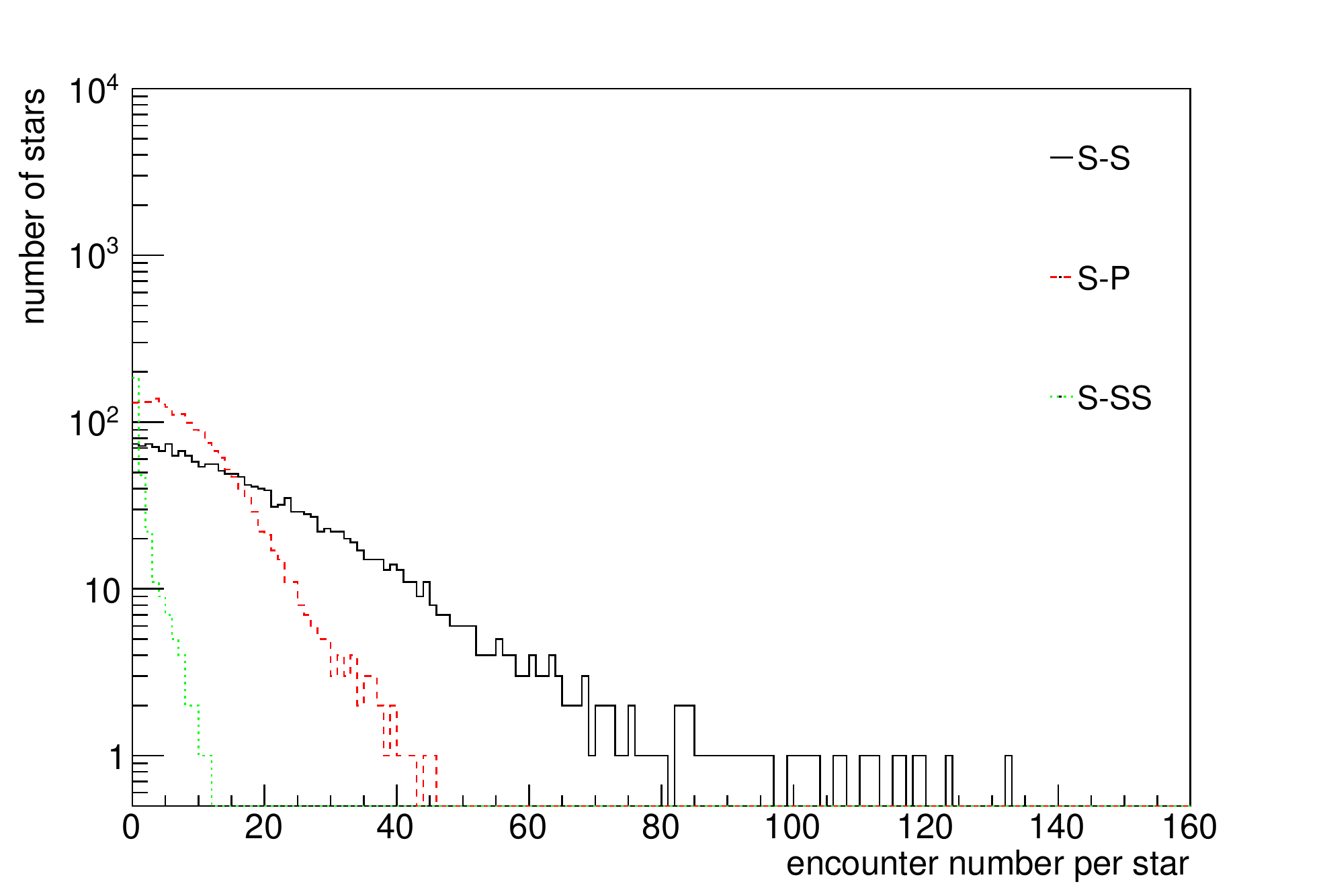}&
	\includegraphics[width=0.5\textwidth]{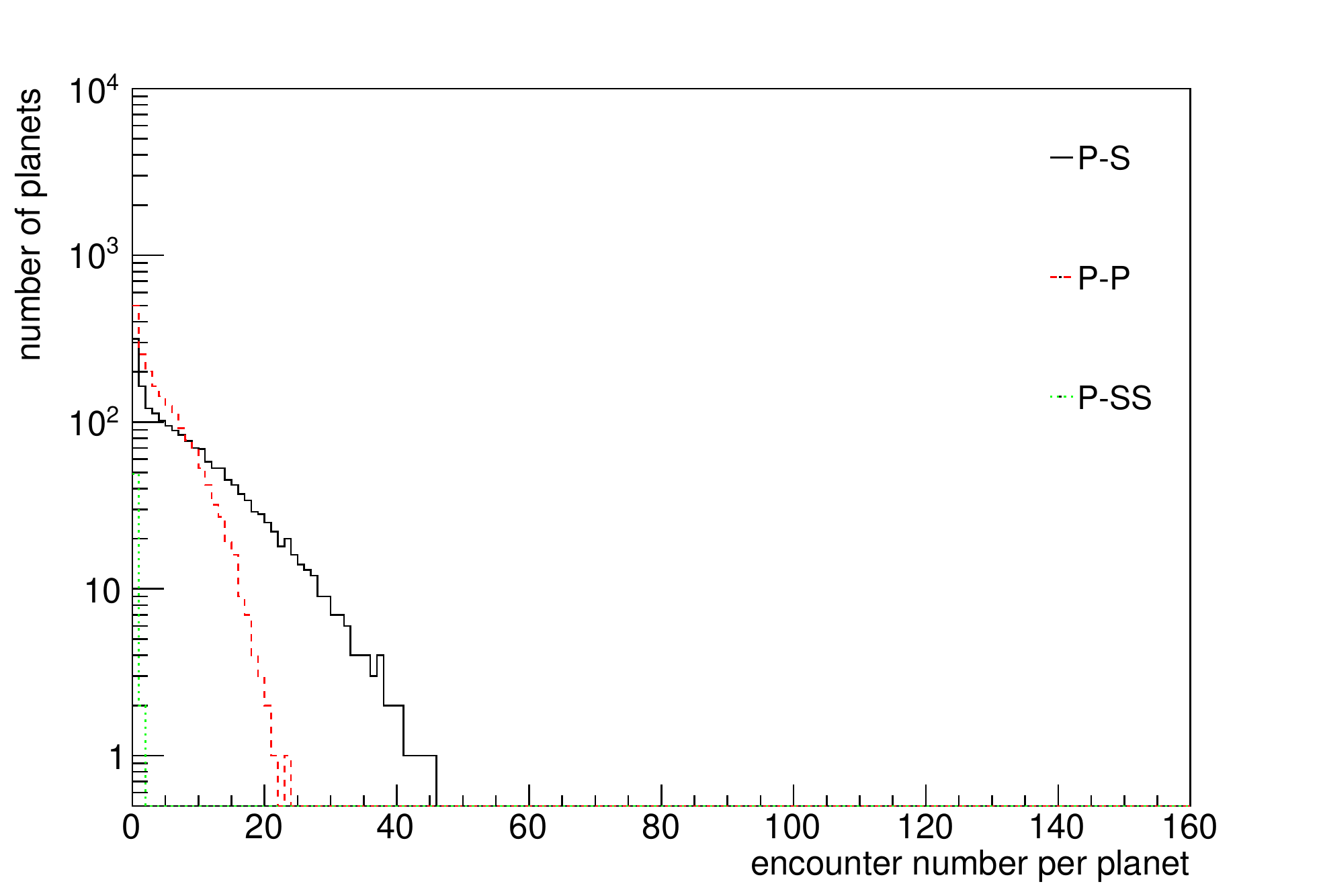}\\
	\includegraphics[width=0.5\textwidth]{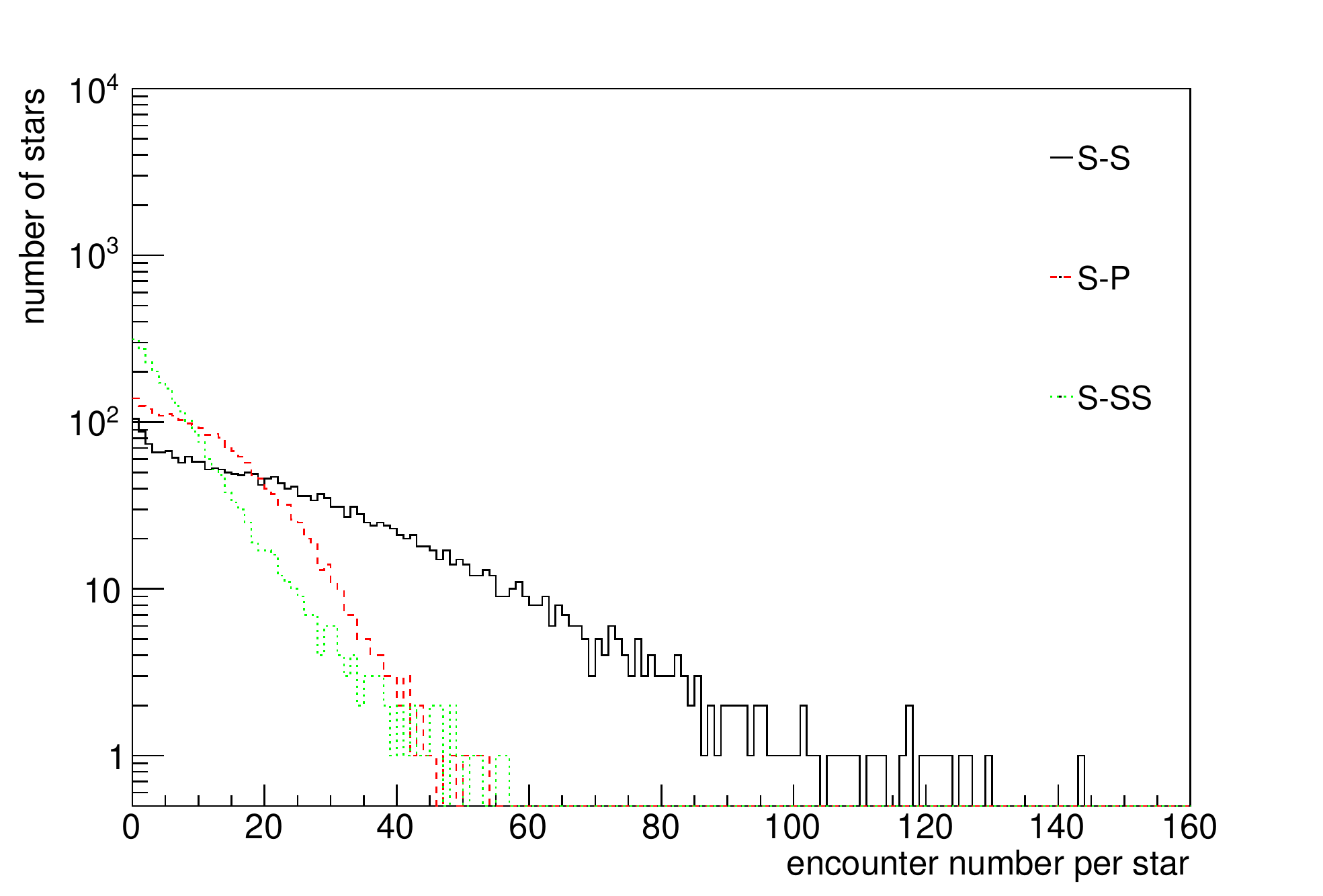}&
	\includegraphics[width=0.5\textwidth]{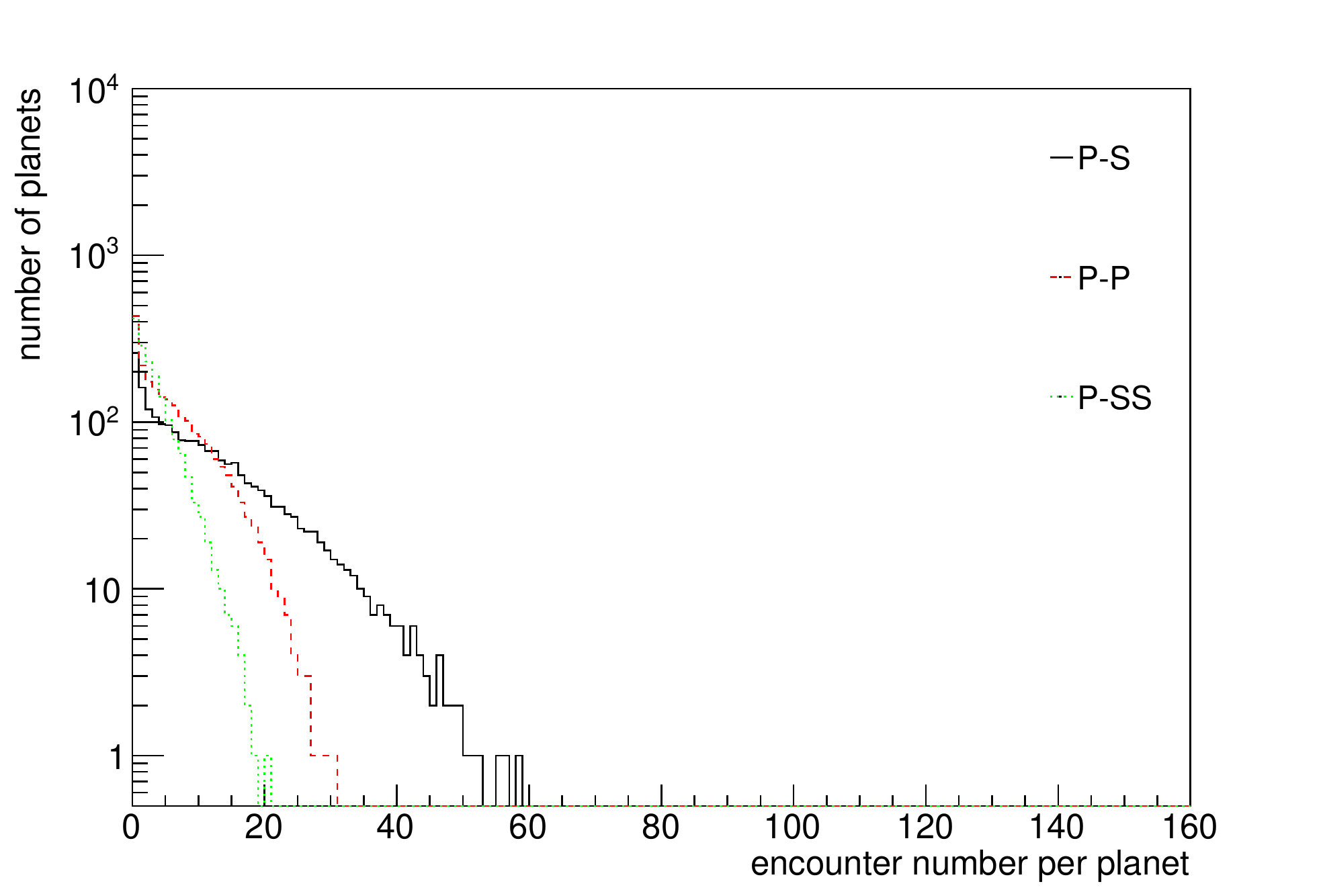}\\
  \end{tabular}
  \caption{The distribution of the total number of close ($<1000$~AU) encounters per star for model~3 ({\em top}) and model~5 ({\em bottom}) during the entire lifetime of the star clusters. Both star cluster models initially contain $N=\ns+\nb=2000$ (single or binary) stellar members with a binary fraction $\binfrac=0\%$ (model~3) or $\binfrac=50\%$ (model~5). Both have an initial planet-to-star ratio $\ratio=2$ and an initial half-mass radius $\rhm=0.38$~pc. Encounter distributions are shown for encounters experienced by single stars ({\em left}) and by FFPs ({\em right}). The different colours indicate different types of encounters. The horizontal axis indicates how many encounters are experienced by a cluster member, and the vertical axis represents the number of cluster members that experience that number of encounters. 
  Each of the histograms represents the average of an ensemble of ten realisations.  
  \label{figure:ecrate} }
\end{figure*}

\begin{figure}
    \includegraphics[width=0.53\textwidth]{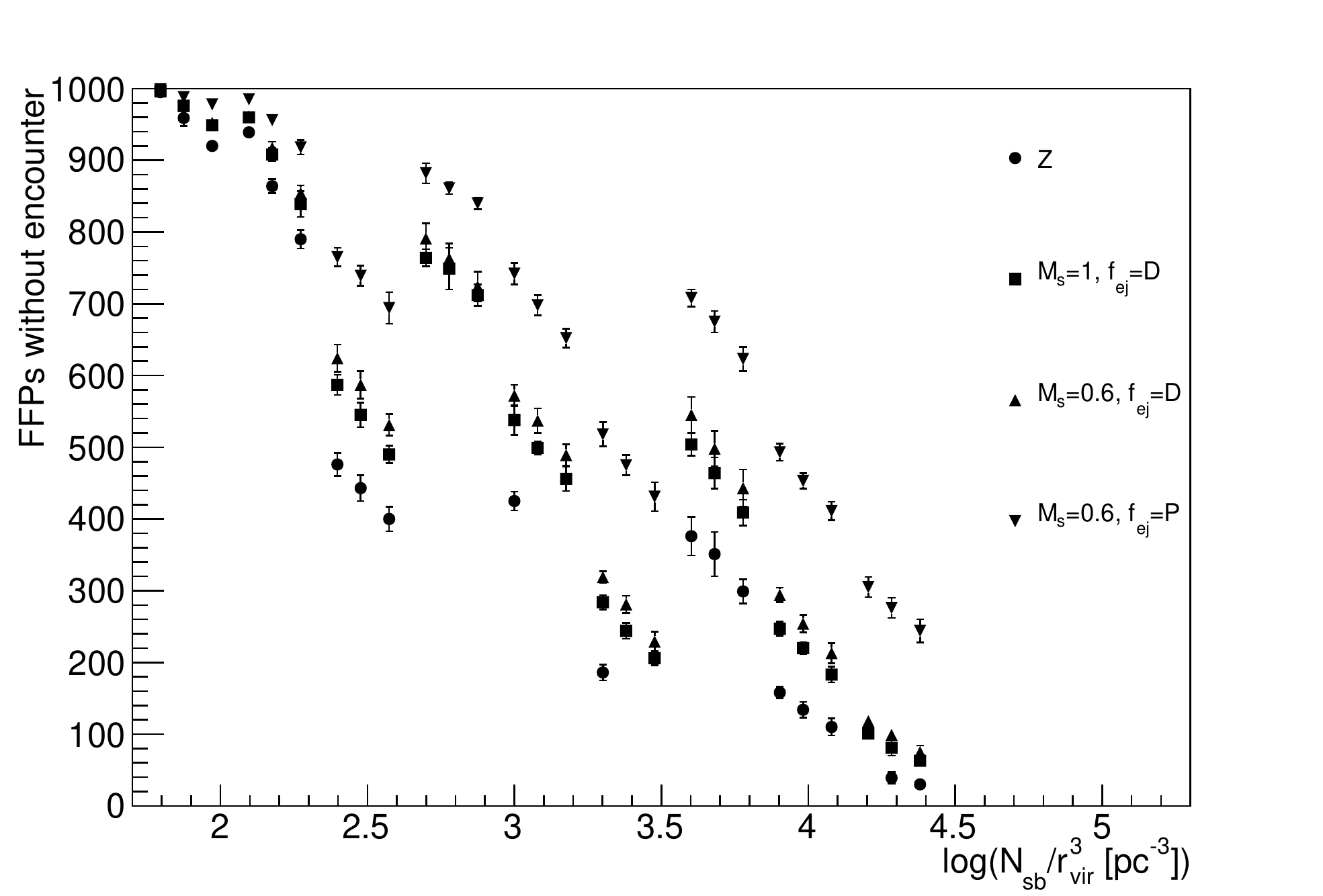}
      \caption{
   The number of FFPs that never approach a star or a binary system within 1000~AU during the entire lifetime of the star cluster. The horizontal axis represents the different models, arranged by the initial number density in the star cluster. Each model initially contains  $\np=1000$ free-floating planets. The adopted initial planet ejection velocity distributions include prompt ejection ($P$), delayed ejection ($D$), and zero ejection velocity ($Z$), and are indicated with the different symbols (see Figure~\ref{figure:ejectionvelocities}). FFPs our lowest-density models ($\nsb=\ns+\nb=500$; $\binfrac=0\%$; $\rvir=2$~pc) almost never encounter a star or binary system, while many FFPs in the highest-density clusters ($\nsb=2000$; $\binfrac=50\%$; $\rvir=0.5$~pc) do approach one or more massive cluster members within 1000~AU. In the latter clusters, the FFPs almost all FFPs experience a close encounter for the model in which the planet ejection velocity is zero ($\velocitytype=Z$).
    \label{figure:ffp_with_massive}}
\end{figure}

In our simulations we also record the total number of close encounters experienced by each individual object. This number depends strongly on the location of the object in the star cluster and on the time it spends in the cluster before escaping. The (combined) total number of encounters experienced by all the objects of type S, P, SS, and SP is:
\begin{equation} \label{eq:totalencounters}
	\begin{array}{ll}
	\nenc_{\rm S,tot}  &=  2\nenc_{S-S} + \nenc_{S-P} + \nenc_{SP-S} + \nenc_{SS-S} \\
	\nenc_{\rm P,tot}  &=  \nenc_{S-P} + 2\nenc_{P-P} + \nenc_{SS-P}\\
	\nenc_{\rm SS,tot} &=  \nenc_{SS-S} + \nenc_{SS-P} + 2\nenc_{SS-SS} \\
	\nenc_{\rm SP,tot} &=  \nenc_{SP-S} \quad\quad \\
	\end{array}
\end{equation}
where we have ignored the encounter types that rarely occur, such as $\nenc_{SP-P}$, $\nenc_{SP-SP}$ and $\nenc_{SS-SP}$ (see also below). Note that during an encounter between two objects of the same type, both objects experience an encounter. Using Eqs.~\ref{eq:encounterrates}, and under the assumption that gravitational focusing is negligible (which may only be a reasonable assumption for encounters involving low-mass bodies), these can be expressed as
\begin{equation} \label{eq:totalencounterssimpler}
	\begin{array}{ll}
	\nenc_{\rm S,tot}  &\propto N^2(1-\binfrac)[2-\binfrac+\np(N^{-1}+1+\binfrac)] \\
	\nenc_{\rm P,tot}  &\propto \np (2\np+N) \\
	\nenc_{\rm SS,tot} &\propto N^2\binfrac(1+\np/N)\\
	\nenc_{\rm SP,tot} &\propto (1-\binfrac^2)N^2\np\\
	\end{array}
\end{equation}
at an instantaneous time $t$, under the assumption that the different populations have the same distribution in phase-space. In the remaining part of this section we discuss the encounters experienced during the entire lifetime of the star clusters, and since all variables in Eq.~\ref{eq:totalencounterssimpler} change with time, these expressions only provide first-order approximations.

The distributions of the number of close encounters per star and per planet, split up into the different components in Eq.~\ref{eq:totalencounters}, are shown in Figure~\ref{figure:ecrate}, for models~3 and~5 (cf. Table~\ref{table:encounterrates}). The left-hand and right-hand panels show the encounters experienced by the stars and by the planets, respectively. Note that in this figure we show the number of encounters experienced by each body (instead of the number of two-body encounters occurring); each encounter corresponds to two individual bodies experiencing an encounter. The area under each histogram corresponds to the values in Table~\ref{table:encounterrates} when the encountering cluster members are of different type (e.g., S-P), and are double the values in Table~\ref{table:encounterrates} when the encountering members are of the same type (e.g., P-P).

The top-left panel of Figure~\ref{figure:ecrate} shows the distribution of the number of encounters experienced by single stars in model~3. Based on our initial conditions, we expect to first order that the total number of encounters of each type is $2\nenc_{S-S} \propto 2\ns$, $\nenc_{S-P}\propto \np$, and $\nenc_{SS-S}\approx 0$, for encounters with other single stars, with FFPs, and with binary systems, respectively. Note that for our choice of the initial conditions, we expect the distributions for the star-star and star-planet encounters to be roughly the same. However, due to the large fraction of FFP escapers at the early stage, the evolution of the star cluster, the formation of new dynamical binaries, and gravitational focusing, the final results differ. Approximately 70~stars (3.5\% of the total number of single stars) experience only one encounter with another star, and a slightly higher number of single stars experiences only one encounter with a FFP. Several single stars, on the other hand, experience over a hundred encounters with other stars, while none of the single stars encounters more than a hundred FFPs. Close encounters of single stars with binary systems are substantially less frequent, as the initial binary fraction is zero and only few binary systems are formed at later times. Approximately 3.5\% of the single stars experience close encounters with one or more dynamically formed binary systems, and most of these occur after the clusters have over half of their stars.

For model~5 (bottom-left panel in Figure~\ref{figure:ecrate}), the number of single stars with a close encounter with one other single star is about 100 (10\%). Although one would expect this number to be lower due to the smaller number of single stars in this model, this value can be explained as a result of the higher mass density in the star cluster, the disruption of wide binaries, particularly in the cluster centre \citep[see, e.g.,][]{grijs2013}, and the longer cluster lifetime, as compared to model~3. Approximately 130 single stars (13\%) experience a close encounter with a FFP. The number of single stars experiencing a close encounter with a binary system is now substantial, due to the initial condition $\binfrac=50\%$. Single stars having many encounters with binary systems occur less frequently compared to single-single encounters, which is partially due to the disruption of the widest binaries over time, and partially due to the fact that binary systems tend to increase the velocities of single stars after an encounter, such that they are more likely to escape the star cluster.

The distribution of close encounters experienced by the planets in models~3 and~5 are shown in the right-hand panels of Figure~\ref{figure:ecrate}. For model~3, approximately 400 (10\%) of the FFPs experience encounters with only one FFP, and a lower fraction (5\%) of the FFPs have only one encounter with a star. All FFPs in the cluster experience fewer than 60 encounters with another star, and fewer than 30 encounters with another FFP. FFPs experience on average more encounters with single stars and with FFPs in model~5 (w.r.t. model~3), since the number density for this model is larger. Again the encounter distributions for encounters with single stars and with FFPs are different. The difference in S-P and P-P encounters between the two models is partially a result of the higher mass concentration, and partially a result of the longer lifetime of model~5. Close encounters between stars and binaries do not occur frequently in model~3 because of the zero initial binary fraction, but as dynamical binaries form, several hundreds of FFPs have a close encounter with a binary system. Encounters of the type SS-P are more frequent in model~5, and the majority of the FFPs have one or more encounters with a binary system. Although the initial binary fraction in model~5 is 50\%, fewer FFPs experience multiple close encounters with binaries than with single stars. 

The encounter distributions for the different types of cluster members in all models have in common that few bodies experience many encounters, while many bodies experience few (or no) encounters. The former are mostly bodies that remain part of the star clusters until it is nearly dissolved, while the latter escape or migrate the the cluster outskirts at early times. This means that statistical averages, such as the results in Eqs.~\ref{eq:totalencounters} and~\ref{eq:totalencounterssimpler} normalised by the number of bodies, should be avoided when estimating, for example, the typical effect of close encounters on existing planetary systems. 

The fraction of FFPs that never experience a close encounter with a stellar-mass particle (single or binary) is shown in Figure~\ref{figure:ffp_with_massive} for the different initial conditions. Note that the initial number density along the horizontal axis is proportional to the initial mass density, as we adopt the same IMF in all models. Thus this result also represent the correlation between FFPs with no encounters and cluster initial mass density. In the lowest-density star clusters none of the FFPs experience a close encounter, while in the higher-density star clusters the fraction depends strongly on the initial velocity distribution of the FFPs. The FFP populations that are initially in virial equilibrium (the filled circles in Figure~\ref{figure:ffp_with_massive}) provide lower limits, and for these populations the fraction of non-encountering FFPs ranges from 0\% to 100\%, primarily depending on the star cluster density.


\subsubsection{Encounter times} \label{section:encountertime}

\begin{figure}
  \centering
  \begin{tabular}{c}
    \includegraphics[width=0.53\textwidth]{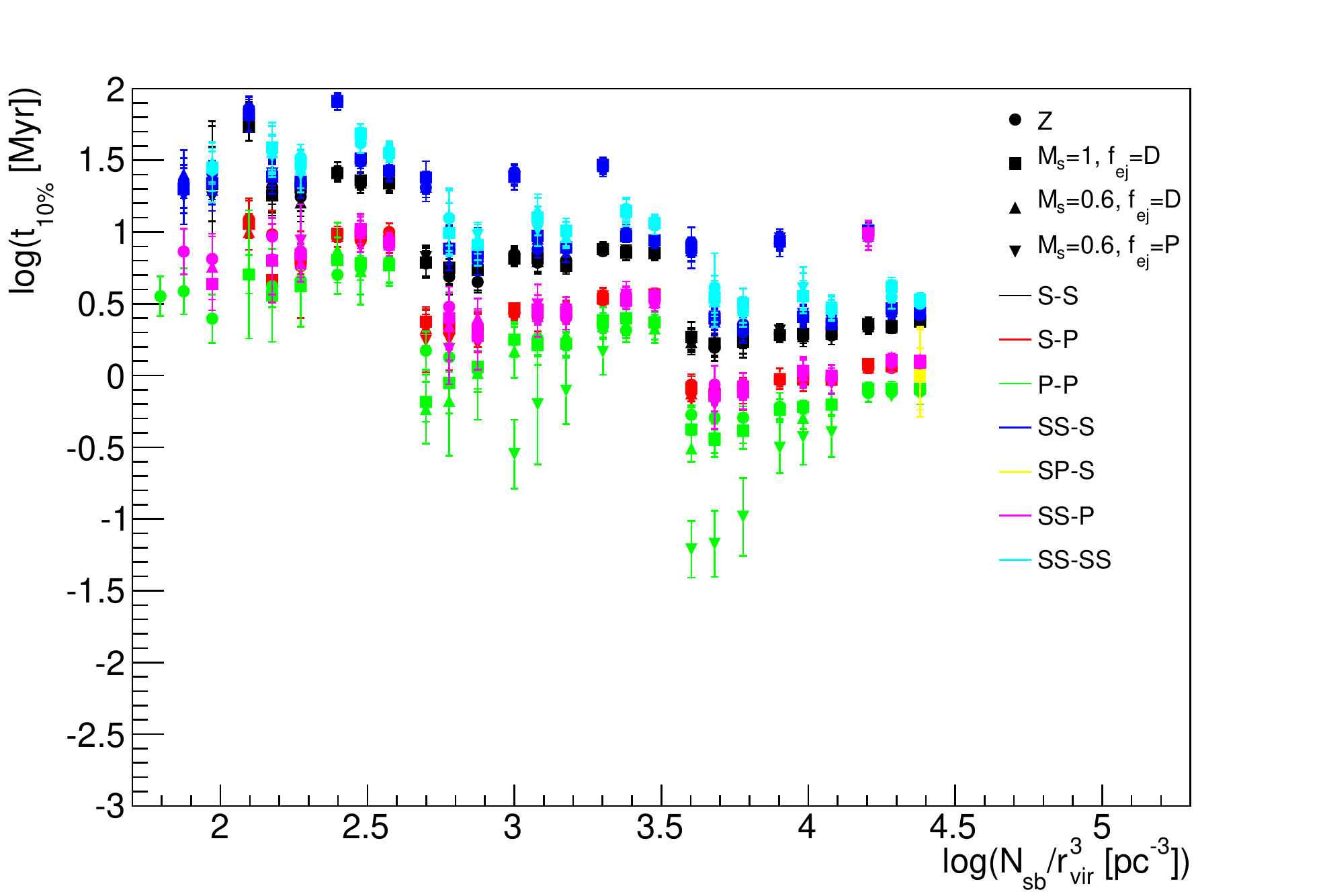}\\	
    \includegraphics[width=0.53\textwidth]{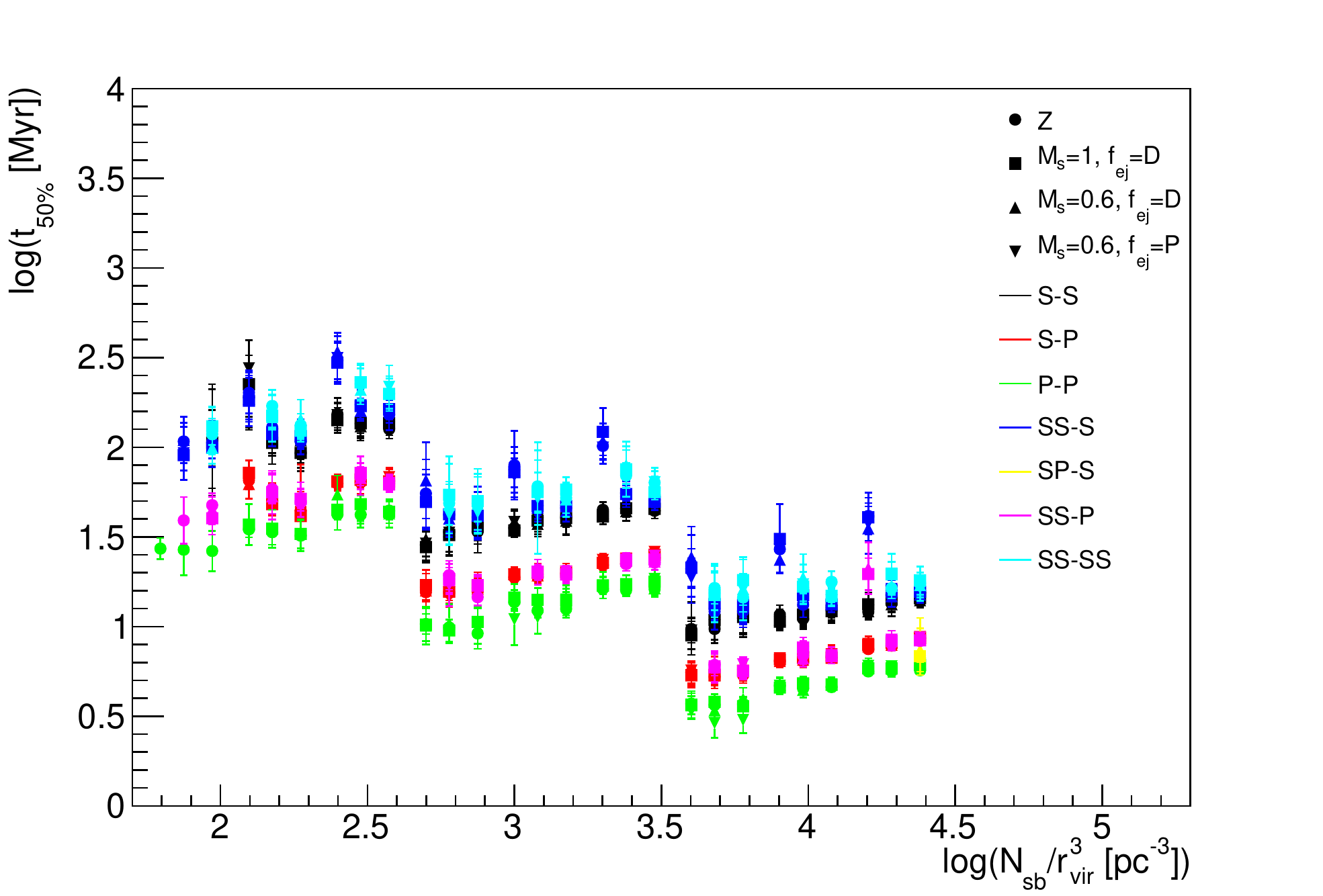}\\
    \includegraphics[width=0.53\textwidth]{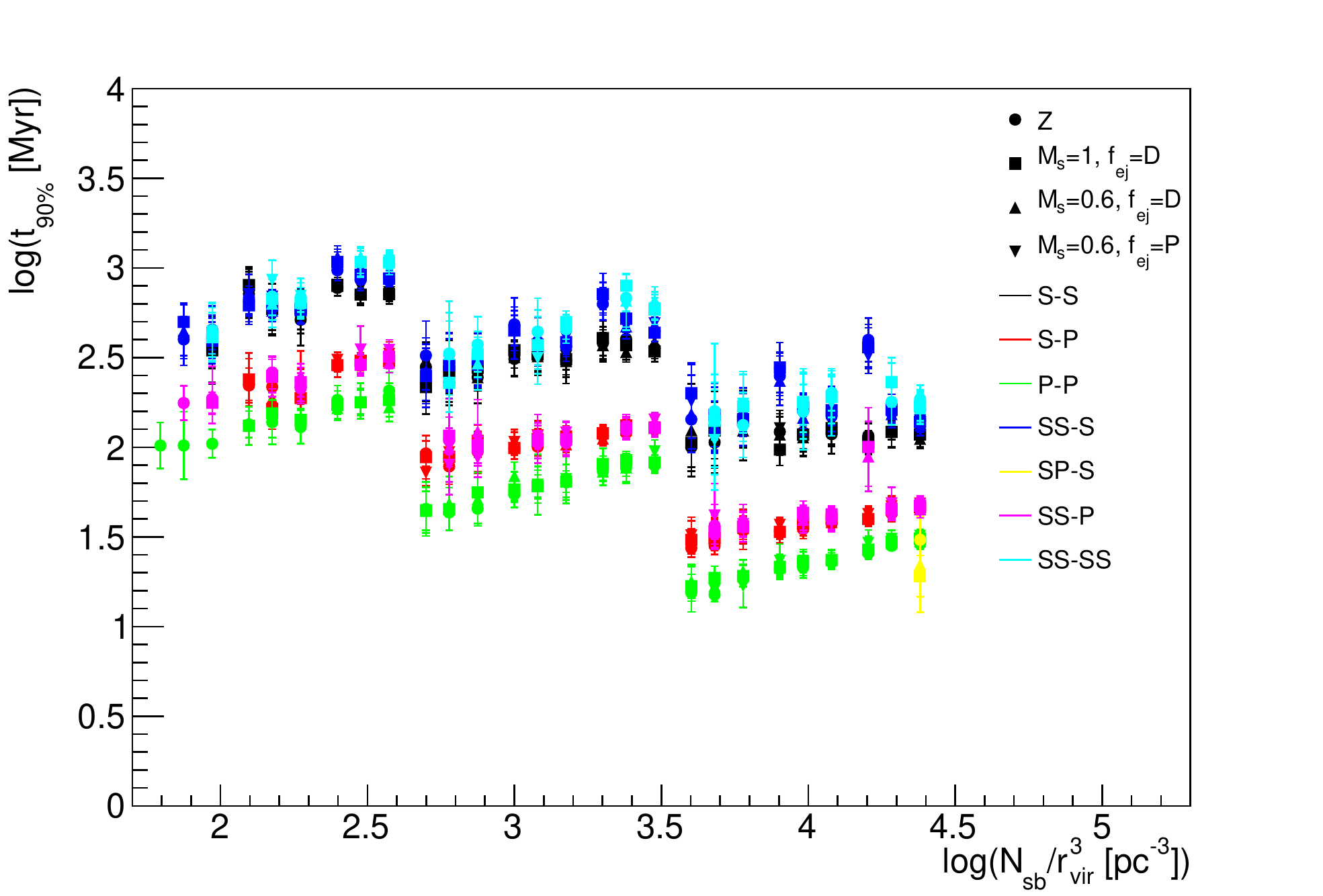}\\
  \end{tabular}
  \caption{
  The times at which 10\% ({\em top}), 50\% ({\em middle}), and 90\% ({\em bottom}) of the encounters have occurred, for all models. The horizontal axis represents the initial number density in the star cluster, from the lowest-density models ($\nsb=\ns+\nb=500$; $\binfrac=0\%$; $\rvir=2$~pc) on the left to the highest-density clusters ($\nsb=2000$; $\binfrac=50\%$; $\rvir=0.5$~pc) on the right. The vertical axis represents the time at which a certain fraction of encounters have occurred (cf. Eq.~\ref{eq:specifichalfencounters}).
  Different colours represent different encounter types between stars (S), FFPs (P), binary systems (SS) and captured planetary systems (SP). The adopted initial planet ejection velocity distributions include prompt ejection ($P$), delayed ejection ($D$), and zero ejection velocity ($Z$), and are indicated with the different markers (see also Figure~\ref{figure:ejectionvelocities}). The data are derived from an ensemble of ten model realisations.
    \label{figure:tfit}}
\end{figure}

The cumulative number of encounters $\nenc_{X-Y}(t)$ between members of two types of populations $X$ and $Y$ should follow
\begin{equation} \label{eq:generalhalfencounters}
	\nenc_{X-Y} \propto \int_0^{\tdiss} N_X(t)N_Y(t)dt
\end{equation}
where $N_X$ and $N_Y$ are the number densities of these populations, respectively. When making the assumption that the proportionality remains constant (i.e., when we ignore mass loss, mass segregation, core collapse, and cluster expansion), the time $t_h$ at which half of the encounters have occurred can be obtained by solving $\nenc_{X-Y}(t_h)=\nenc_{X-Y}(\tdiss)/2$ for $t_h$. Approximate values for $t_h$ for the various types of encounters can be derived by substituting Eqs.~\ref{eq:starloss} and/or~\ref{eq:solution} into Eq.~\ref{eq:generalhalfencounters}, giving
\begin{equation} \label{eq:specifichalfencounters}
	\begin{array}{lll}
		t_{h,S-S} & \approx (1-2^{-1/3})\tdiss & \approx 0.21\tdiss \\
		t_{h,S-P} & \approx (1-2^{-1/4})\tdiss & \approx 0.16\tdiss \\
		t_{h,P-P} & \approx (1-2^{-1/5})\tdiss & \approx 0.13\tdiss\\
	\end{array}
\end{equation}
where we have for simplicity assumed that the FFP population is initially virialized, that $\binfrac=0$, and that Eq.~\ref{eq:starloss} holds . The expressions can be generalized for the cases with $\np(t_1)<\np(0)$ and $t_1>0$ using Eq.~\ref{eq:ratiostarloss}, and will result in smaller values of $\nenc(t_h)$ for the encounters involving planets. Note that Eq.~\ref{eq:specifichalfencounters} is independent of the number of stars and the number of planets, if the assumptions stated above hold.

The timescales at which $10\%$, $50\%$ and $90\%$ of the encounters of the specific types occur, are shown in Figure~\ref{figure:tfit} for all modelled star clusters. Note that we model each star cluster until it has completely dissolved. The times at which these events occur vary with the initial properties of the star clusters and their FFP populations, and also with the types of encounters. In general, however, $10\%$ of the encounters occur in the first 3~Myr, $50\%$ occur within 30~Myr, and 90\% of the encounters occur within 100~Myr.

The middle panel shows the times at which half of the encounters have occurred, and its values can be approximated with Eq.~\ref{eq:specifichalfencounters}. Although $\tdiss$ depends on the star cluster properties (particularly on $\rvir$, $\nsb$ and $\binfrac$), the ratios $t_{h,S-S}/t_{h,S-P}\approx 1.31$ and $t_{h,S-S}/t_{h,P-P}\approx 1.62$ are independent of these, which explains why the differences in $t_h$ in Figure~\ref{figure:tfit} are roughly constant in logarithmic units. The real differences in the figure are slightly larger than in Eq.~\ref{eq:specifichalfencounters} because several of the assumptions are violated, as a result of expansion and early loss of FFPs in the systems. This can be observed in the top panel, which shows that the 10\% encounter time of the P-P interactions is substantially smaller than the simple approximation suggests for the models with high planet ejection velocity. For these cases, a better expression may be derived by not adopting $\np(t_1)=\np(0)$ and $t_1=0$ in Eq.~\ref{eq:solution}, as done above. However, even though a large number of planets may escape at early times, the majority of the P-P encounters occur between the FFPs that remain in the star clusters for a long time, and therefore the results in the middle and bottom panels depend only mildly on $\velocitytype$. The encounters involving massive components (such as SS-S and SS-SS) generally occur at later times than those with low-mass components, as the former remain part of the star cluster for longer times. Figure~\ref{figure:tfit} also demonstrates that encounters occur at later times for systems with increasing $\nsb$, $\binfrac$, or $\rvir$, since these systems have larger relaxation times (Eq.~\ref{eq:relaxationtime}) and generally have longer dissolution times.


\subsection{Encounter velocities and periastron distances}


\subsubsection{Encounter velocity distributions}

\begin{figure*}
  \begin{tabular}{cc}
    \includegraphics[width=0.5\textwidth]{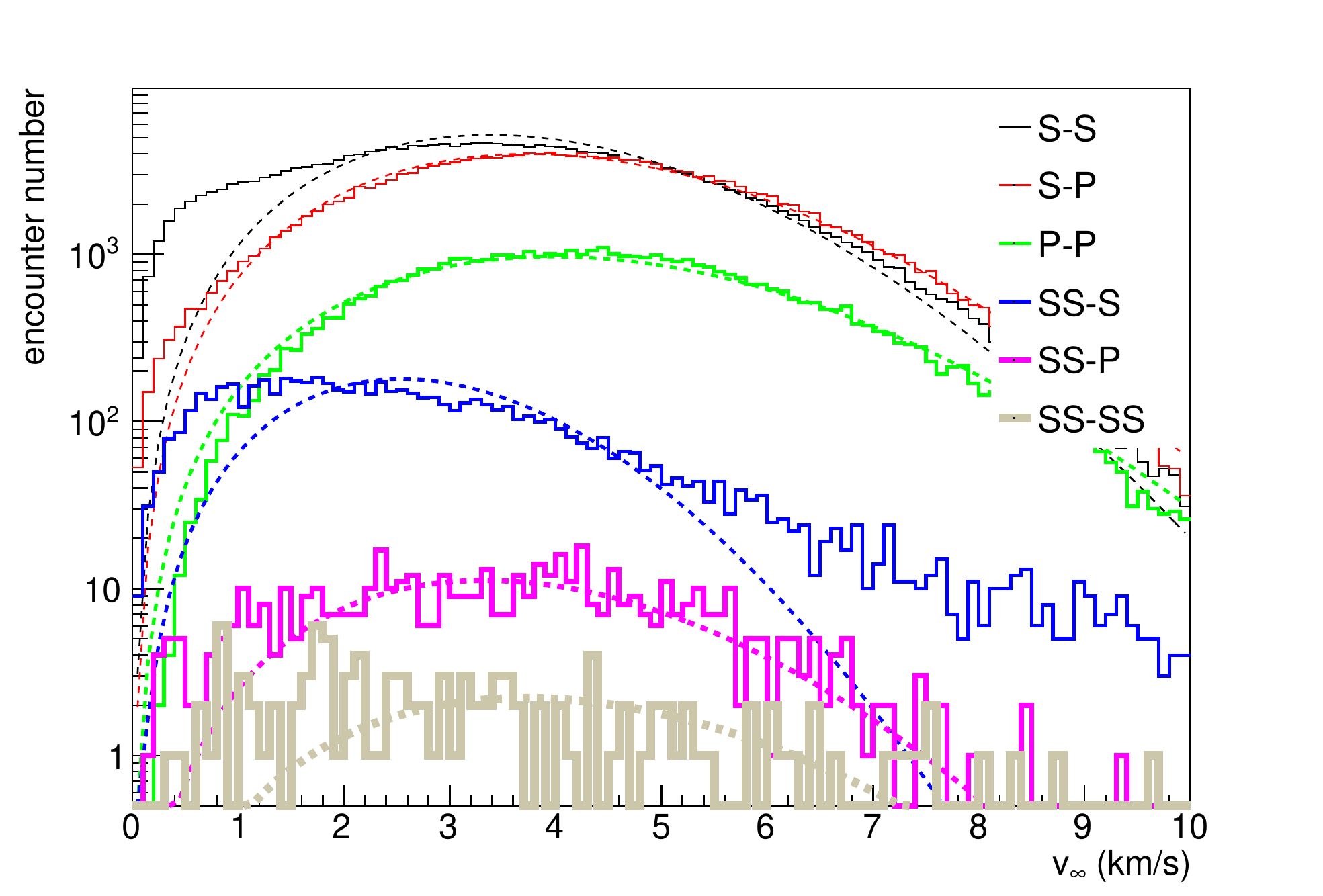} &
    \includegraphics[width=0.5\textwidth]{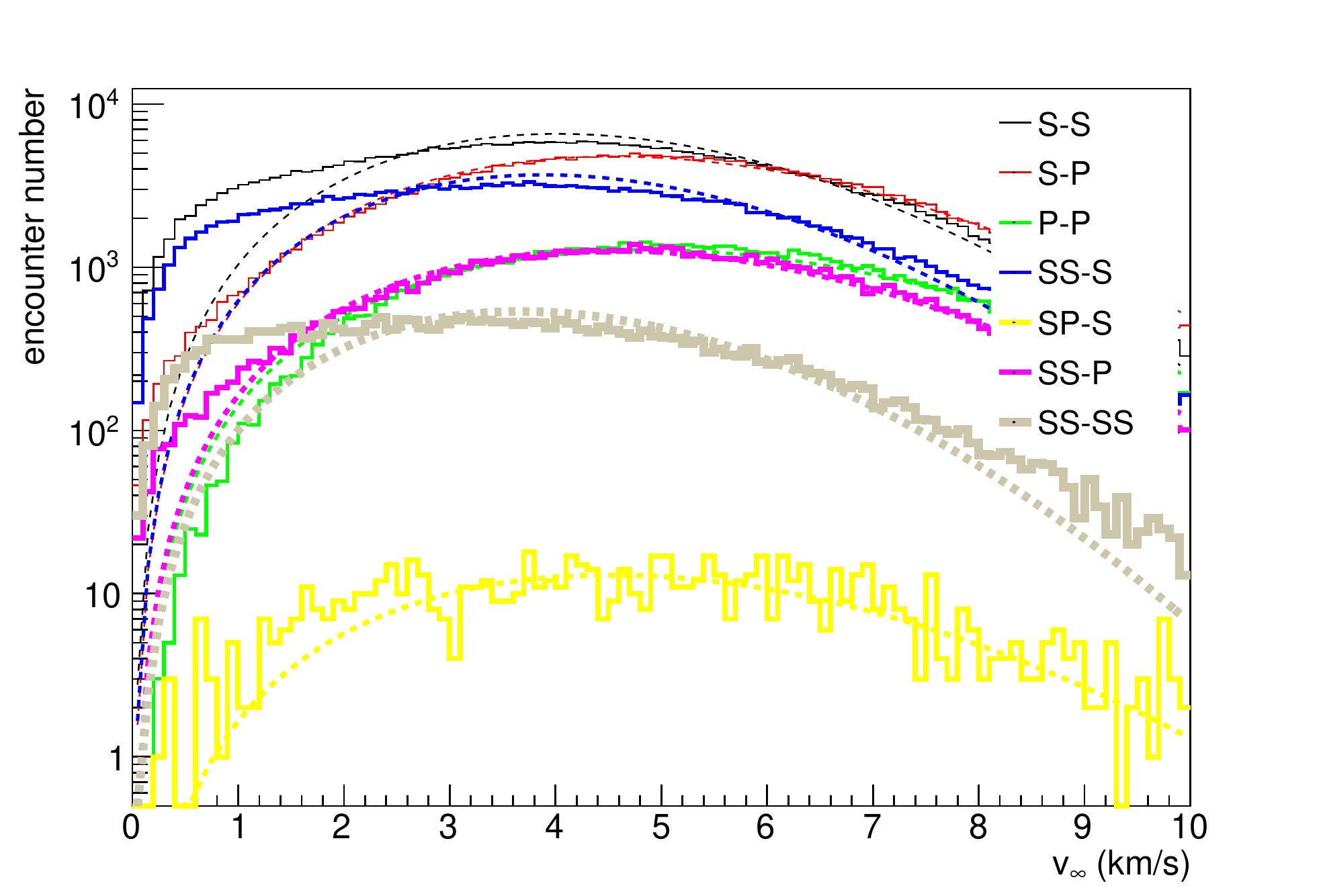}\\
    (a) $N=2000$, $\binfrac=0\%$~$\ratio=2$, $\rhm=0.38$~pc&
    (b) $N=2000$, $\binfrac=50\%$~$\ratio=2$, $\rhm=0.38$~pc\\
    \includegraphics[width=0.5\textwidth]{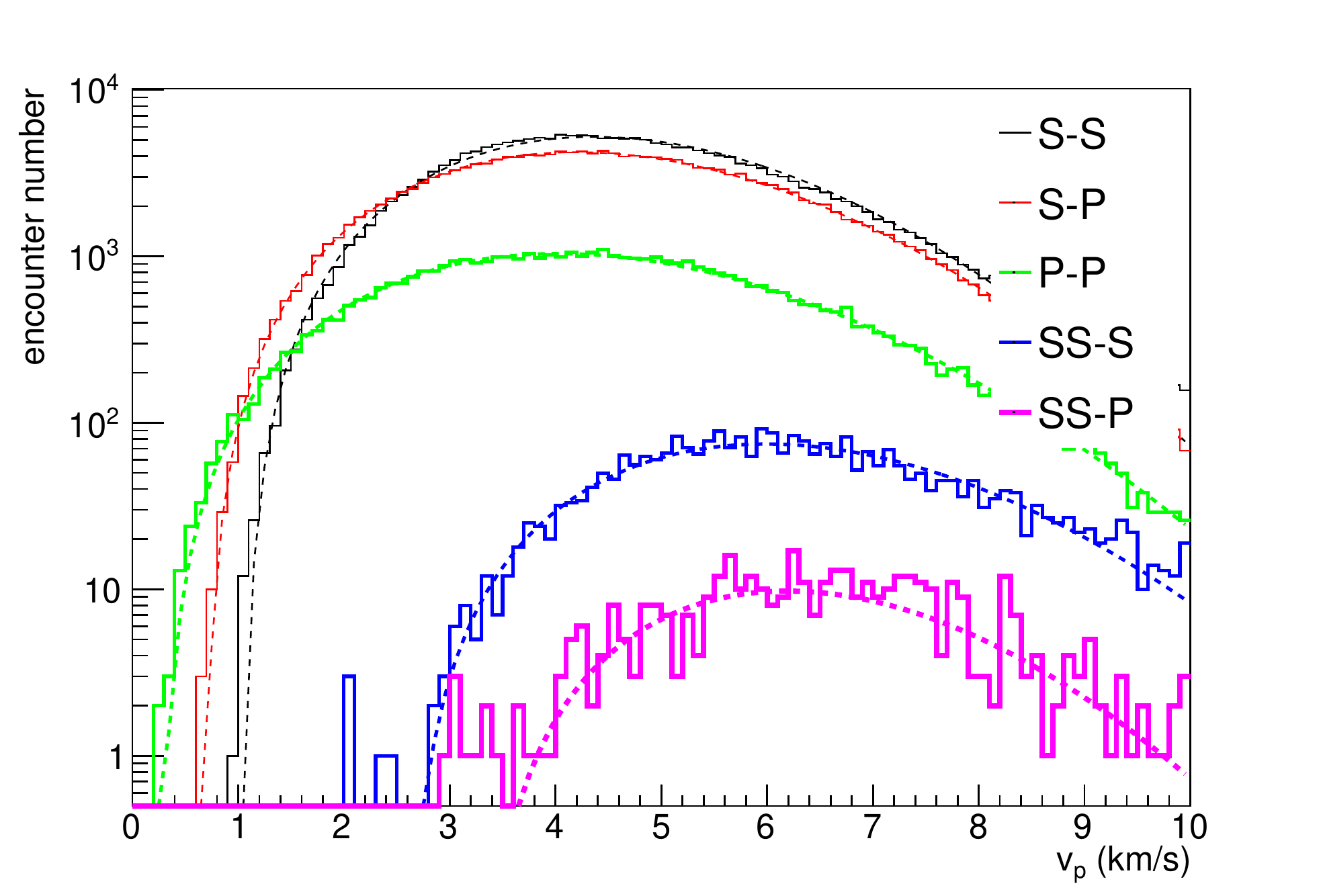} &
    \includegraphics[width=0.5\textwidth]{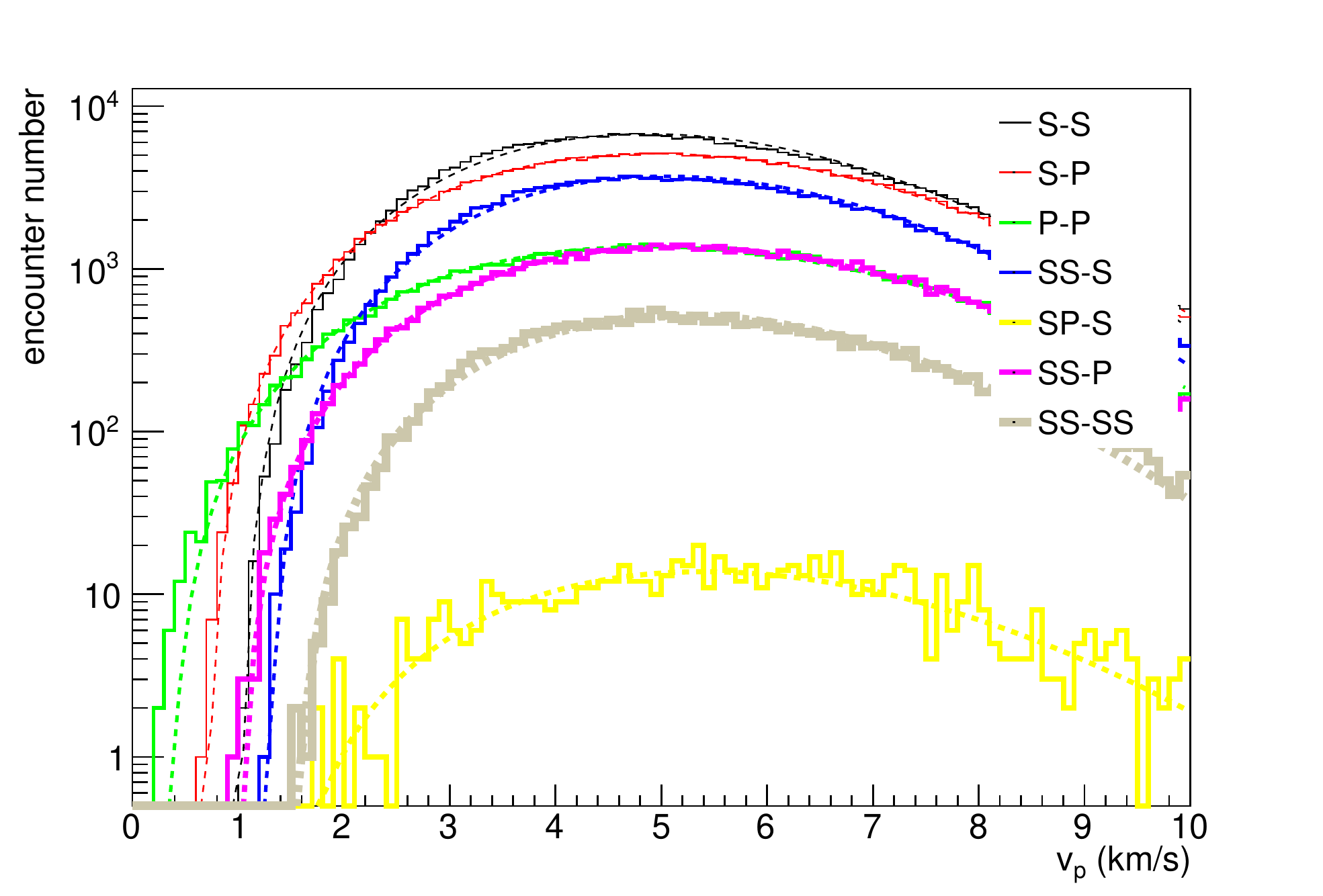}\\
    (c) $N=2000$, $\binfrac=0\%$, $\ratio=2$, $\rhm=0.38$~pc &
    (d) $N=2000$, $\binfrac=50\%$, $\ratio=2$, $\rhm=0.38$~pc\\
    \end{tabular}
  \caption{The velocity-at-infinity ($\vinf$) distributions of the close encounters, for model~3 ({\em top-left}) and model~5 ({\em top-right}), and the periastron velocity ($\vperiastron$) distributions of the close encounters for model~3 ({\em bottom-left}) and model~5 ({\em bottom-right}). 
  The histograms represent the simulation results, and the curves are best fits to Eq.~\ref{eq:maxwellian} for $\vinf$ and to Eq.~\ref{eq:maxwellianoffset} for $\vperiastron$.  
  The data represent the combined results for ten realisations.  
     \label{figure:incomingvelocity} }
\end{figure*}

The trajectories of two approaching unbound bodies in isolation are hyperbolic, and have an associated velocity-at-infinity $\vinf$. The velocity-at-infinity distribution, $f(\vinf)$, of an ensemble of two-body encounters in a star cluster is approximately a Maxwell-Boltzmann distribution:
\begin{equation} \label{eq:maxwellian}
  	f(\vinf) 
	= \left(\pi \vpmax^2\right)^{-3/2}
	\exp \left(-\frac{\vinf^2}{\vpmax^2} \right) 4\pi\vinf^2 
\end{equation}
where $\vpmax$ is the most probable value of $\vinf$. The corresponding mean value is $\langle \vinf \rangle = 2\pi^{-1/2}\vpmax$. Eq.~\ref{eq:maxwellian} is generally a good approximation and is independent of mass of the bodies involved in the encounter, provided that the cluster is not mass segregated and that the star cluster's velocity distribution is isotropic. For an encounter with impact parameter $b$, the velocity at periastron can be expressed as
\begin{equation}
	\vperiastron = \vinf \sqrt{ \frac{e+1}{e-1} } 
	 \quad\quad {\rm with} \quad\quad
     e = \sqrt{1+\left(\frac{b\vinf^2}{GM}\right)^2} \ ,
\end{equation}
where $e$ is the hyperbolic eccentricity. The distribution of impact parameters follows $f(b)\propto b$. When the combined mass of the bodies is small, such as for P-P encounters, the equation reduces to $\vperiastron=\vinf$. For a close, low-$\vinf$ encounter between massive bodies, the expression reduces to $\vperiastron\approx 2GM/b\vinf$. 
We approximate the distribution of the velocities at periastron, $f(\vperiastron)$ with a modified Maxwell-Boltzmann distribution that has a velocity offset $\vpoffset$ and a dispersion parameter $\vpmax$:
\begin{equation} \label{eq:maxwellianoffset}
  	f(\vperiastron) 
	\propto \left(\pi \vpmax^2\right)^{-3/2}
	\exp \left(-\frac{(\vperiastron-\vpoffset)^2}{\vpmax^2} \right) 4\pi(\vperiastron-\vpoffset)^2 \ ,
\end{equation}
where $\vpoffset=\vperiastron-\vinf > 0$ is the velocity increase during the encounter and $\vpmax$ is the most probable periastron velocity.

The histograms in Figure~\ref{figure:incomingvelocity} show the distributions of $\vinf$ ({\em top}) and $\vperiastron$ ({\em bottom}) for the various types of encounters that occur in star cluster models~3 ({\em left-hand panels}) and~5 ({\em right-hand panels}), and the corresponding best fits to Eqs.~\ref{eq:maxwellian} and~\ref{eq:maxwellianoffset}. The curves represent the distributions integrated over all locations in the star clusters, for their entire evolution. Since the encounter velocities depend both on the location within the star cluster and on time, Eqs.~\ref{eq:maxwellian} and~\ref{eq:maxwellianoffset} are approximations. Nevertheless, since most encounters occur within the first $\sim 100-200$~Myr, the approximations are reasonable.

The strongest deviations from the Maxwellian distribution are seen for the largest values of $\vinf$ and $\vperiastron$, indicating interactions in the centre of the star cluster, and for the smallest values of $\vinf$ and $\vperiastron$, indicating encounters in the cluster outskirts and during the latest phases of the evolution when the stellar density is low. 
Since planetary systems have $\vperiastron\approx\vinf$, the corresponding curves for the P-P encounters are very similar in all four panels, and well-fitted by Eq.~\ref{eq:maxwellian}. All encounters involving planets (P-P, S-P, SS-P, and P-P) are qualitatively very similar, both for model~3 and model~5. As a result of gravitational focusing, the typical velocity at periastron increases as the total mass of the encounters increases.

With respect to the Maxwellian fit, excess S-S interactions with small $\vinf$ occur. These interactions are most frequent in the low-density regions in the outskirts of the clusters and also at late stages of evolution, when the effect of gravitational focusing on low-velocity stars is more prominent. The deviation from a Maxwellian distribution is much stronger for the SS-S interactions. Since model~3 does not have primordial binaries, all binaries involved in the SS-S encounters are formed through capture. As these binaries are generally formed in the cluster centre, and as their components are generally of high mass, Eq.~\ref{eq:maxwellian} does not provide a good approximation for the distribution. Since $\binfrac=50\%$ for model~5, the effect of gravitational focusing and longevity of the binary population is now much better visible. The SS-S and SS-SS show deviations from the Maxwell-Bolzmann distributions, because of gravitational focusing and because they tend to remain longer in the star cluster. 
Note that the dynamically-formed binary systems in model~3 have on average substantially larger $\vperiastron$ than the primordial binaries in model~5, although they are fewer in number. As these dynamical binaries in model~3 are generally more massive than the primordial binaries in model~5, their encounters with other stars and FFPs are stronger, and therefore their periastron velocities are larger.

There are about ten times more SP-S encounters in model~5 than in model~3, indicating that the presence of binary stars in star clusters enhances the star-planet captures. The distribution $f(\vinf)$ for SP-S encounters is a scaled-down version of that of the S-S encounters, which may be expected since the planetary companions do not play a role in the encounter trajectory or encounter rates.

\begin{figure}
  \centering
  \includegraphics[width=0.53\textwidth]{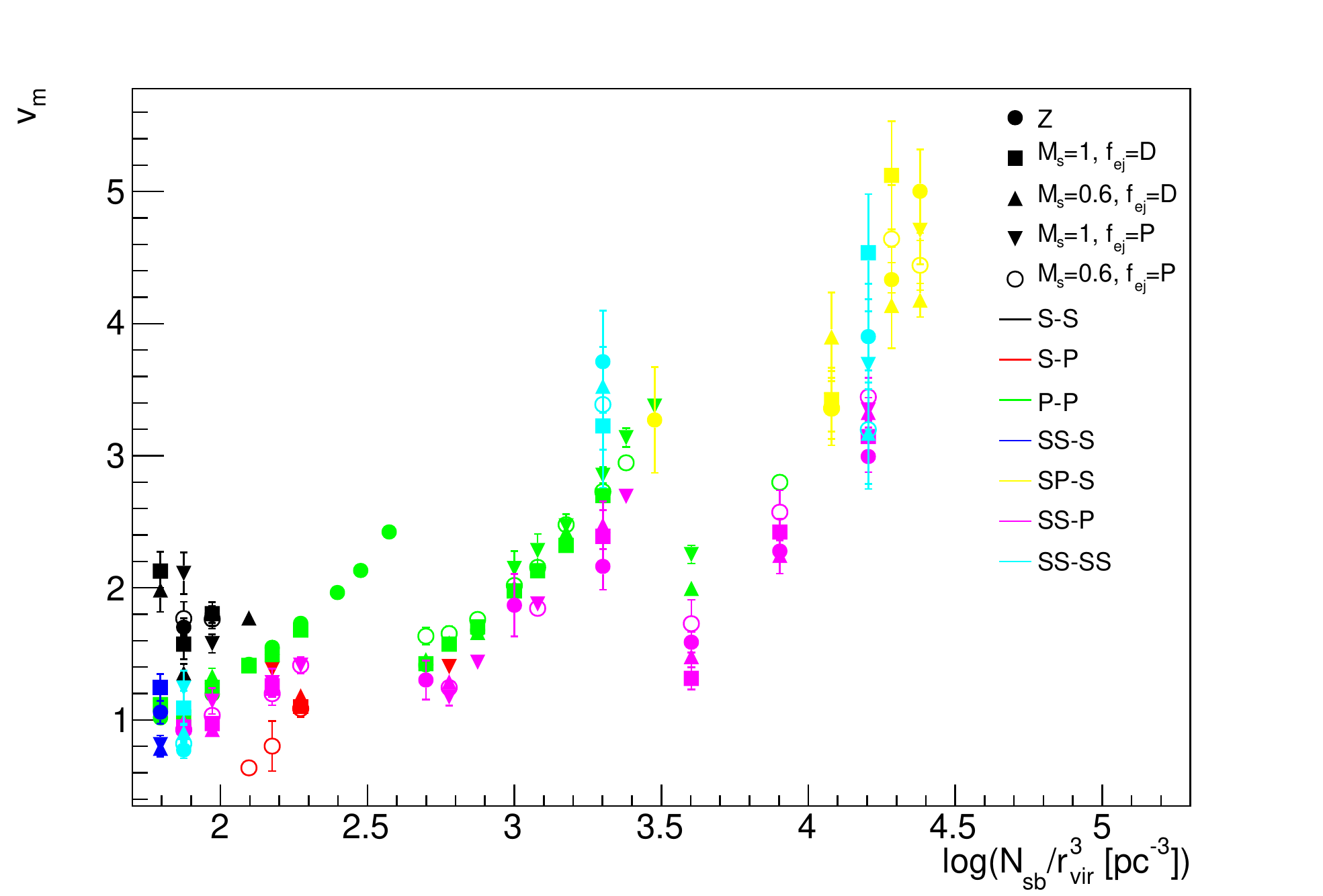}\\
  \includegraphics[width=0.53\textwidth]{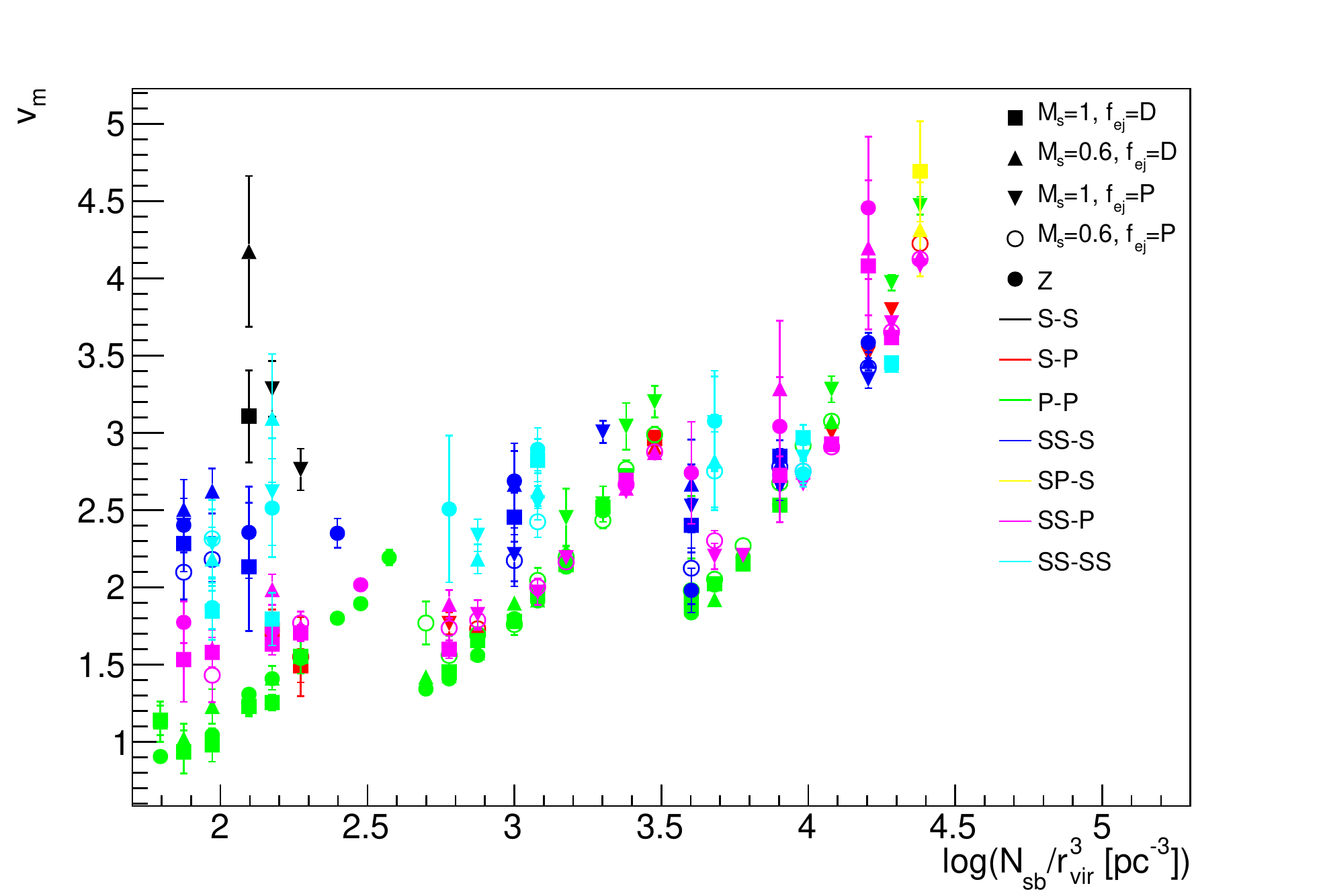}\\
  \includegraphics[width=0.53\textwidth]{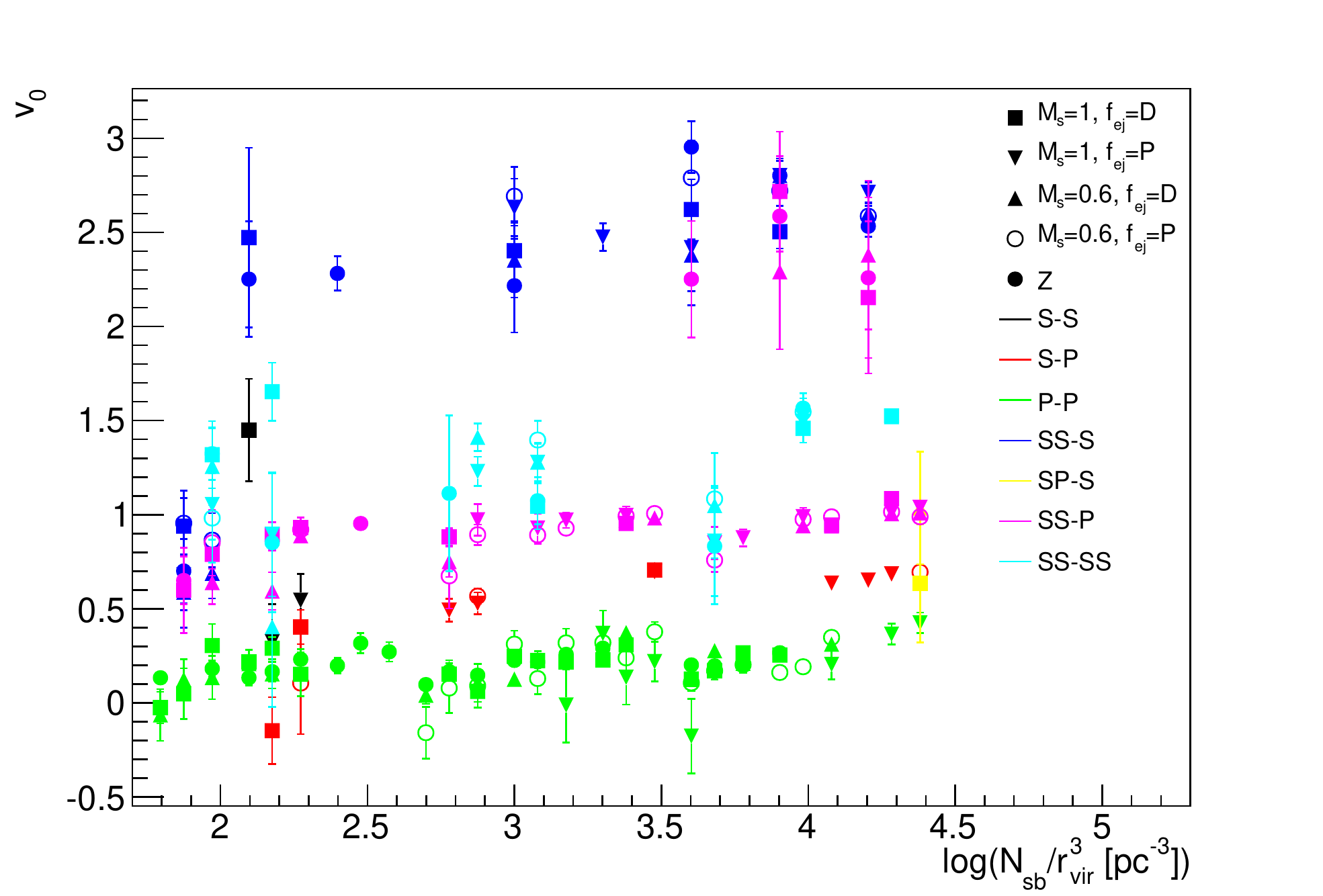}\\[-2ex]
  \caption{Properties of the velocity-at-infinity distributions $f(\vinf)$ and periastron velocity distributions $f(\vperiastron)$ for the different types of encounters for the modelled set of star clusters. The horizontal axis represents the initial stellar density, as in Figure~\ref{figure:tfit}. {\em Top:} the fitted parameter $\vpmax$ (Eq.~\ref{eq:maxwellian}) to the measured distribution $f(\vinf)$. The two other panels show the best-fitting parameters $\vpmax$ ({\em middle}) and $\vpoffset$ ({\em bottom}) in Eq.~\ref{eq:maxwellianoffset} for the measured distribution $f(\vperiastron)$. 
  All models resemble model~3 (cf. Figure~\ref{figure:incomingvelocity}). Different colours represent different encounter types. Data with fitting errors larger than 0.5 are omitted. Different markers indicate the initial planet ejection velocity distributions, including prompt ejection ($P$), delayed ejection ($D$), and zero ejection velocity ($Z$); cf. Figure~\ref{figure:ejectionvelocities}). Data are derived from an ensemble of ten model realisations.\label{figure:vpfit}}
\end{figure}

The best-fitting values of $\vpmax$ and $\vpoffset$ in Eqs.~\ref{eq:maxwellian} and~\ref{eq:maxwellianoffset} are plotted in Figure~\ref{figure:vpfit} for the different types of encounters, for all simulated models. Note, however, that we did not include the bad fits, which result either from distributions that are substantially different from Eq.~\ref{eq:maxwellian}, or from low-number statistics.

Most encounters occur during the early phase of star cluster evolution when mass segregation is small, and during this period we expect $\vpmax$ to be directly proportional to the velocity dispersion of the star cluster (Eq.~\ref{eq:velocitydispersion}). For star cluster members near the half-mass radius $\vpmax \propto \sigma(r=\rhm) \propto (\nsb/\rhm)^{1/2}$. This trend is clearly seen for most fits to $f(\vinf)$ plotted in Figure~\ref{figure:vpfit}, and is independent of $\velocitytype$, $\np$, and the type of encounters. Since the P-P encounters have $\vperiastron \approx \vinf$, we obtain the same relation for $f(\vperiastron)$. For the encounters between massive components the fitted values for $\vpmax$ are larger, and $\vpmax$ typically increases with increasing component masses.

Since FFPs barely change their velocities during P-P encounters, we expect $\vpoffset\approx 0$. The scatter for the P-P interactions results from the fact that we have fitted the distributions $f(\vinf)$ and $f(\vperiastron)$ to all encounters integrated over the entire volume of the star cluster and its entire evolution. The value of $\vpoffset$ depends only mildly on $\binfrac$, $\nsb$, $\velocitytype$ and $\rhm$. As the fitted value $\vpmax$ takes into account the dependency of $f(\vperiastron)$ on $\vinf$, the value of $\vpoffset$ is primarily determined by the mass of the encountering components. Since gravitational focusing is stronger for encounters between massive bodies, the fitted value $\vpoffset=\vperiastron-\vinf$ increases with the combined mass of the members involved in the encounter.

\begin{figure}
  \centering
  \includegraphics[width=0.53\textwidth]{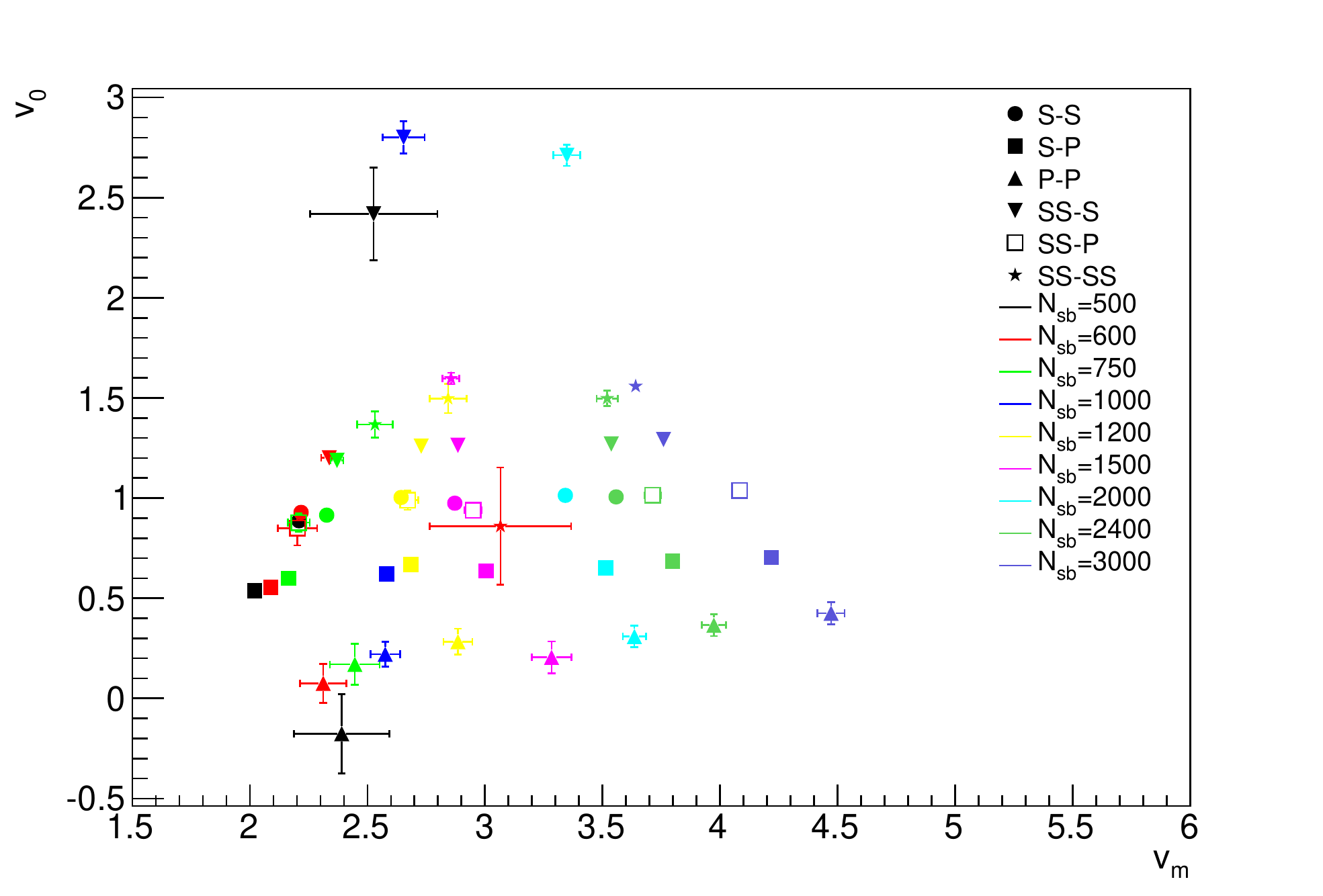}\\
  \caption{Properties of the periastron velocity distributions $f(\vperiastron)$ for the different types of encounters for the modelled set of star clusters. The fitted parameters $\vpmax$ and $\vpoffset$ to Eq.~\ref{eq:maxwellianoffset} are indicated along the horizontal and vertical axes, respectively, for the models with $\rhm=0.38$~pc, $\np=1000$ and with ejection velocity distributions corresponding to the prompt-$1\msun$ distribution in Figure~\ref{figure:ejectionvelocities}. The different markers indicate the different types of encounters between stars (S), FFPs (P), binary systems (SS), and captured planetary systems (SP). The values of $\nsb=\ns+\nb$ are indicated with the different colors. \label{figure:vpvcfit}}
\end{figure}

The fitting results for $f(\vperiastron)$ are combined in Figure~\ref{figure:vpvcfit}. The value of $\vpmax$ is primarily determined by the velocity dispersion of the star cluster, while $\vpoffset$ is mostly determined by combined mass of the encountering bodies. These trends are seen in Figure~\ref{figure:vpvcfit}: models with identical velocity dispersion lie roughly on vertical lines, while the same types of encounters (S-S, S-P, etc) lie roughly on a horizontal line. Apart from some scatter, the values of $\vpoffset$ of most of the encounter types indeed lie roughly on a horizontal line in  Figure~\ref{figure:vpfit}, although there is a trend of $\vpoffset$ increasing with $\vpmax$ (and therefore with increasing $\nsb$). This trend is strongest for the P-P encounters. 
The results in Figure~\ref{figure:vpvcfit} can be used to estimate the periastron velocity distributions for other star cluster masses and for encounters between members with other masses, using Eq.~\ref{eq:maxwellianoffset}. It can also be used to statistically estimate the probability of obtaining a certain periastron velocity for a single free-floating planet in a star cluster.

\subsubsection{Encounter periastron distance distribution} \label{section:encounterperiastron}

\begin{figure}
  \centering
    \includegraphics[width=0.53\textwidth]{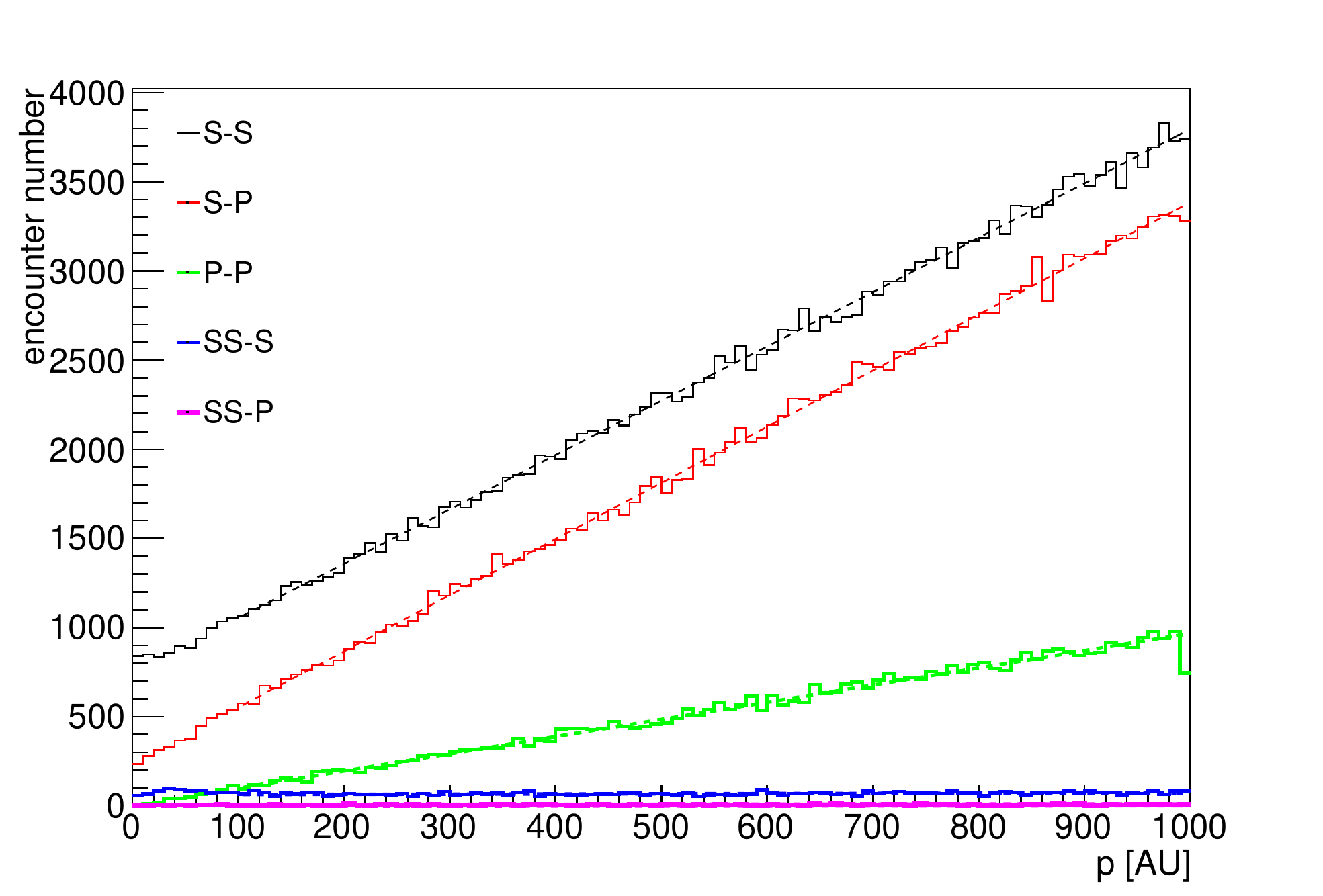}\\[-47ex]
    \quad\quad\quad\quad(a) $N=2000$, $\binfrac=0\%$, $\ratio=2$, $\rhm=0.38$~pc\\[45ex]
    \includegraphics[width=0.53\textwidth]{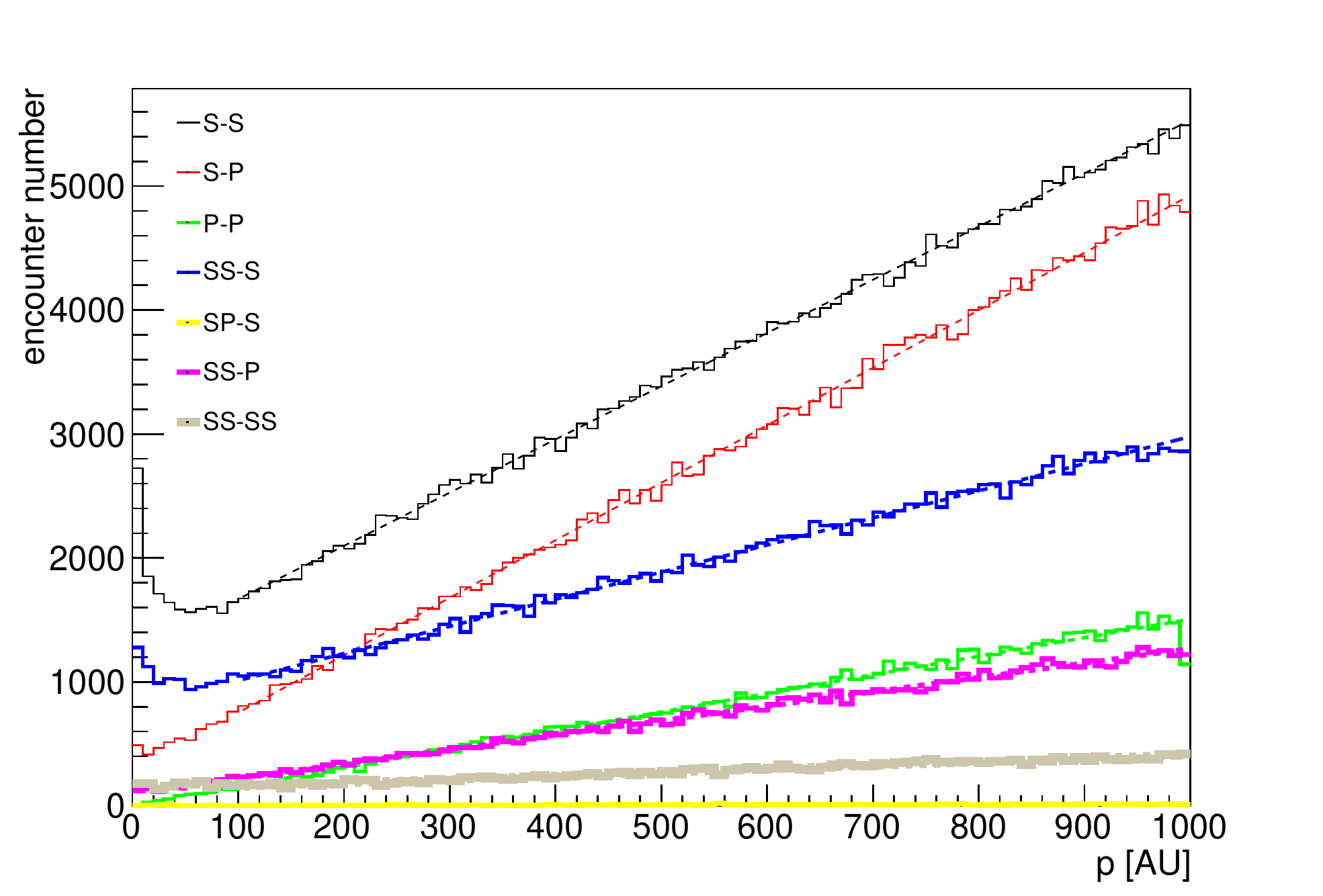}\\[-47ex]
    \quad\quad\quad\quad(b) $N=2000$, $\binfrac=50\%$, $\ratio=2$, $\rhm=0.38$~pc \\[45ex]
  \caption{The periastron distance distributions $f(p)$ of the close
    encounters for model~3 ({\em top}) and model~5 ({\em bottom}). Different colours represent different types of encounters between stars (S), FFPs (P), binary systems (SS) and captured planetary systems (SP). The solid histograms indicate the number of close encounters in each bin, and the dashed lines are linear fits. Encounter types that occur less than 200 times (e.g., SS-SS) are omitted.  The curves represent the combined results for ten realisations. \label{figure:periastron} }
\end{figure}

\begin{figure}
  \centering
  \includegraphics[width=0.53\textwidth]{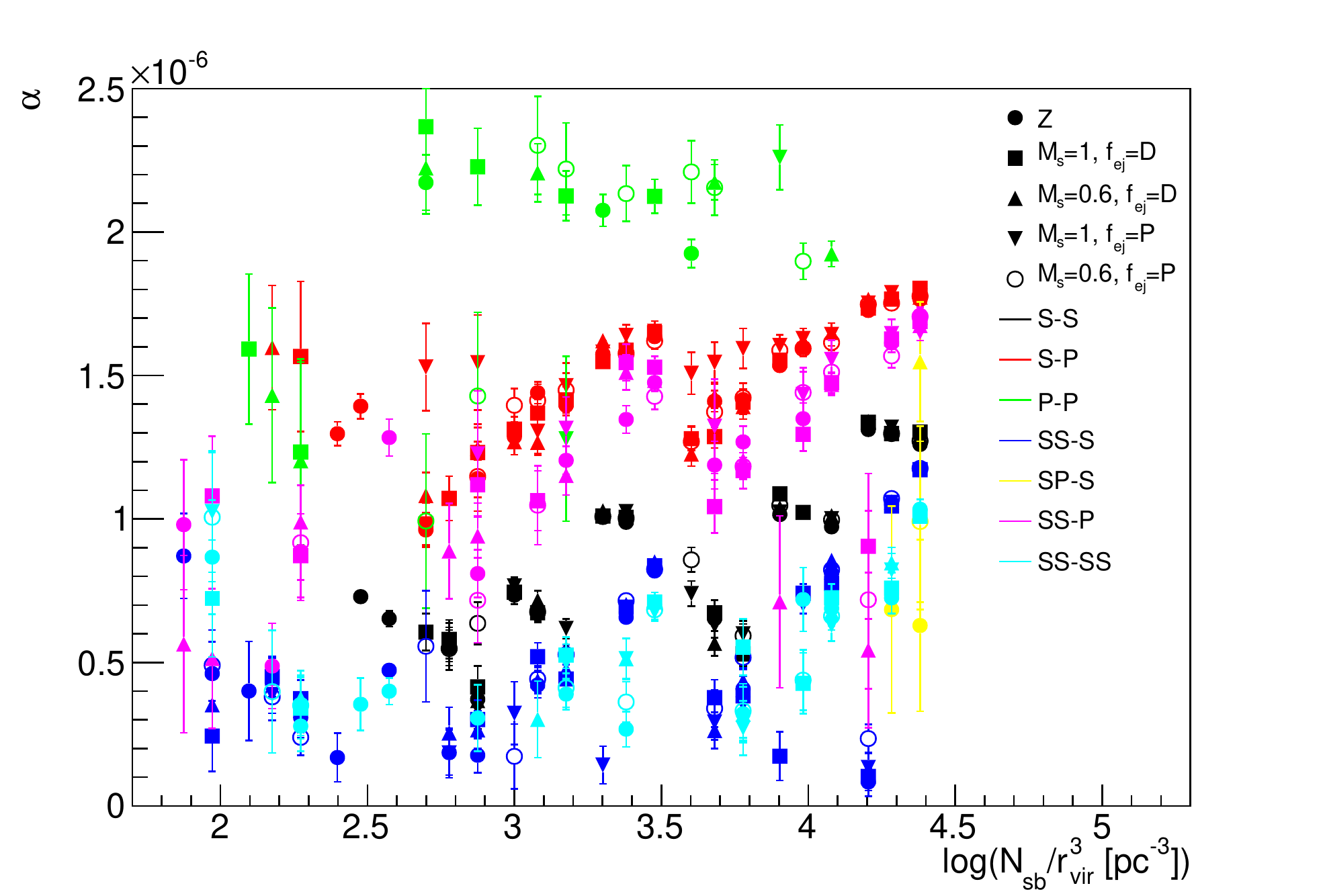}\\
  \caption{The fitted proportionality parameter $\alpha$ (Eq.~\ref{eq:periastrondistribution}) for the periastron distance distributions $f(p)$ for the different star clusters.  The horizontal axis represents the initial stellar density, as in Figure~\ref{figure:tfit}. The adopted initial planet ejection velocity distributions are indicated with the different markers, and include prompt ejection ($P$), delayed ejection ($D$), and zero ejection velocity ($Z$); see Figure~\ref{figure:ejectionvelocities}. The different types of encounters between stars (S), FFPs (P), binary systems (SS) and captured planetary systems (SP) are indicated with the colors. Results are shown for a combined ensemble of ten realisations of each model. \label{figure:pfit}}
\end{figure}

For hyperbolic orbits between two objects with masses $M_1$ and $M_2$, the
relation between the impact parameter $b$, the periastron distance $\periastron$, and
the velocity at infinity $\vinf$ is given by:
\begin{equation}
  b = \periastron \sqrt{1 + \frac{2GM}{\periastron \vinf^2} } \ , 
\end{equation}
where $M=M_1+M_2$ is the combined mass of the two objects. For two approaching bodies in a virialized star cluster, the velocity at infinity $\vinf$ is typically the
velocity dispersion in the star cluster multiplied by a factor $\sqrt{2}$. When both bodies have a negligible mass, such as in the case of a planet-planet encounter, the equation reduces to $\periastron=b$, while for a close encounter between two large masses the effect of gravitational focusing is important and the equation reduces to $\periastron = (b\vinf)^2(2GM)^{-1}$. As the distribution over impact parameters scales as $f_b(b)\propto b$, the corresponding distribution for $p$ is $f(\periastron)\propto \periastron$ for distant encounters. The corresponding distribution for close encounters can be obtained with transformation of variables, which results in $f(\periastron) \propto \vinf^2(2GM)^{-1}$. In other words, $f(\periastron)$ is independent of $p$ in this regime.

Figure~\ref{figure:periastron} shows the periastron distance distributions $f(\periastron)dp$ for the different types of encounters for models~3 and~5. The distributions are linear over most of the range in $p$, and we therefore describe the results using the functional form
\begin{equation} \label{eq:periastrondistribution}
	f(\periastron) = \nenc \left( \alpha \periastron + \beta \right)
	= \nenc \left( \alpha (p-p_m/2) + p_m^{-1}\right)
\end{equation}
to the distributions for the different types of encounters, where $\nenc$ represents the total number of encounters of a specific type, $\alpha$ is in units of AU$^{-2}$ and $\beta$ in units of AU$^{-1}$. Normalization to the total number encounters gives $\beta=p_m^{-1}-\alpha p_m/2$, where $p_m=1000$~AU is the largest periastron distance that is taken into consideration. In the limit that the masses of the encountering bodies are negligible, $f(p)=2\nenc p/p_m^2$, while in the limit of very close encounters between massive bodies, $f(\periastron)=\beta=\nenc/p_m$. The corresponding cumulative distribution is
\begin{equation}
	F(\periastron) = \nenc \left( \frac{\alpha p(p-p_m)}{2}  + \frac{p}{p_m} \right) \ ,
\end{equation}
which indicates the cumulative number of encounters smaller than $\periastron$.
Note that in this approximation the values of $\alpha$ (and $\beta$) are independent of the number of encounters, and also independent of $\nsb$, $\np$, and $\binfrac$. For encounters between more massive bodies, however, $\alpha$ (and therefore also $\beta$) depend on $\vinf$, and the latter quantity may have a different distribution in space (due to mass segregation) and time (due to the preferred escape of low-mass members) than for P-P encounters.

In realistic star clusters $f(\periastron)$ is a combination of weak, intermediate, and strong encounters, and for the combination of all encounters of a certain type, integrated over the entire star cluster and over time. Nevertheless, $f(\periastron)$ is still reasonably well approximated by the expression in Eq.~\ref{eq:periastrondistribution}. Several of the distributions in Figure~\ref{figure:periastron} show an overabundance of periastron approaches below 10~AU with respect to Eq.~\ref{eq:periastrondistribution}. This overabundance is exclusively seen for encounters involving single stars (S-S, S-P, and S-SS) in model~5, and is related to the evolution of primordial binary systems, their interactions with other star cluster members, and subsequent decay.

Figure~\ref{figure:pfit} shows the fitted values of $\alpha$ (Eq.~\ref{eq:periastrondistribution}) for all models.
Since the vast majority of the P-P encounters are weak, the fitted value for $\alpha$ is close to the predicted value in the unfocused limit, $\alpha=2p_m^{-2}=2\times 10^{-6}$~AU$^{-2}$, although the scatter around this value is considerable, particularly for the low-density star clusters for which a large fraction of the FFPs escape immediately. Larger masses for the encountering bodies generally result in smaller values of $\alpha$, which is consistent with the increased gravitational focusing. In addition, when the star cluster mass increases and/or its radius decreases, the velocity dispersion (and hence $\vinf$) becomes larger, such that gravitational focusing is less important and the measured values of $\alpha$ are larger.


The typical time that two encountering members spend near periastron is $\Delta t = 2\periastron/\vperiastron$. For weak encounters this is $\Delta t \approx 2b/\vinf$ and for strong encounters this is $\Delta t \approx b\vinf/GM$.
This timescale is important in studying the effects of encounters when one of the two members, for example a single star, has a planetary system, as it describes, given a planet orbiting a star with a certain semi-major axis, whether the encounter is impulsive or adiabatic, and whether it is close or tidal \citep[see the extensive study of][for further analysis]{spurzem2009}.


\section{Conclusions and Discussion} \label{section:conclusions}

Free-floating planets (FFPs) may be abundant in young star clusters, as close encounters between stars can destabilise planetary systems, which can result in direct or delayed ejection of planets from their host planetary systems. In this article we have presented a study of the dynamical evolution of FFP populations in various types of star clusters. We have carried out $N$-body simulations to characterise how the FFP population evolves and we have studied the properties of close ($<1000$~AU) encounters between single stars, binary stars, and FFPs. Our results can be summarised as follows:
\begin{enumerate}

\item A certain fraction of the FFPs escape from their star clusters shortly after they are ejected from their host stars, while the remaining FFPs escape at much later times. Beyond the time of early escape, the planet-to-star ratio decreases linearly with time, and the times $\thalfstar$ and $\thalfplanet$ at which respectively half of the stars and planets escape from the star clusters are related through the simple relation $\thalfplanet\approx 0.6\thalfstar$. For individual planets with low ejection velocities this means that they are likely to escape from the star cluster at a 40\% earlier time than their host star.

\item Many FFPs ejected from their host star system at early times experience tens of close ($<1000$~AU) encounters with other stars and FFPs in the star cluster before escaping from the cluster. The fraction of FFPs that leave the cluster without any close encounter increases with increasing initial velocity distribution and with decreasing stellar density.

\item Typically half of the encounters of all types occur within 30~Myr, while only 10\% of the encounters occur after the first 100~Myr. The ratios of the timescales at which half of the number of star-star, star-planet, and planet-planet encounters occur are $t_{h,S-S}:t_{h,S-P}:t_{h,P-P}\approx1:0.77:0.62$.

\item The velocity-at-infinity distributions, $f(\vinf)$, of the encounters are well approximated with Maxwell-Boltzmann distributions. The periastron velocity distributions for all types of encounters are well fitted by distributions $f(\vperiastron-\vpoffset)$ similar to a Maxwell-Boltzmann distribution, with a velocity offset $\vpoffset$. The most frequent velocity $\vpmax$ in this distribution is proportional to the velocity dispersion of the star cluster, and the offset $\vpoffset$ is primarily determined by the combined mass of the encountering bodies of a certain type.

\item The distribution over periastron distances is linear, $f(\periastron)=\alpha p + \beta$ for the encounters is linear over most of the values of $p$. In our case we have recorded all approaches within $p_m=1000$~AU. In that case, distant encounters have $\alpha=2\times 10^{-6}$~AU$^{-2}$ and $\beta=0$~AU$^{-1}$. In the case of close encounters, $f(\periastron)=\beta=\nenc/p_m$, where $\nenc$ is the total number of encounters and $p_m$ is the close encounter limit. Also when we combine the encounters for the entire life of the star clusters, the linear fit is good, except for very small distances ($p<10$~AU), for which approaches are more common.

\end{enumerate}
Our study is intended to obtain a general overview of the dynamical evolution of a FFP population in star clusters. We have made several assumptions that should be kept in mind when making realistic predictions for existing star clusters, or when interpreting observational data. First, a critical assumption we made is that the FFP space density is identical to the stellar density. In reality, the ejection velocities of FFPs may depend in an intricate way on the position and velocity of the star at the moment when it is perturbed. This is especially the case when planets are ejected promptly after their host system experiences a close encounter. One would then expect planets to be preferably ejected when their host star is near the star cluster centre. In the case of delayed ejection, however, this may well be a good approximation, since the stars are well-mixed at the moment of planet ejection. 

Second, in our initial model setup we have also made the crude assumption that all planets are initially free-floating. We also neglect any additional FFPs produced by the close encounters between stars in our model, as well as the possible capture of FFPs by planetary systems. Although these is an unrealistic assumptions, particularly in the case of perturbed planetary systems that may eject planets tens of millions of years later \citep[e.g.,][]{malmberg2011, hao2013}, it is a good approximation if most planet ejections occur at early times. Moreover, many of the analytical and computational results presented in Section~\ref{section:results} are easily scalable to any value of $\np$. These can therefore also be used to evaluate the probabilities for close encounters and escape for individual planets ($\np=1$). When convolved with an time-dependent FFP production rate, estimates for the dynamical behaviour of the entire population of FFP can subsequently be obtained.

In a follow-up study we will analyse more realistic initial conditions, ideally by modelling the full $N$-body evolution of decaying multi-planet systems in young star clusters, potentially using the AMUSE framework \citep[][]{portegies2013, pelupessy2013, cai2015}, 
Specifically, we will analyse the dynamical evolution of a FFP population where the FFPs have a stronger preference to be generated in the cluster centre, and where they are ejected at appropriate times resulting from close encounters with other cluster members and the decay of perturbed multi-planet systems.


\section*{Acknowledgments}

We wish to thank Sourav Chatterjee for carefully reading the manuscript and providing useful comments that helped to improve this paper.
We wish to thank Rainer Spurzem and Sverre Aarseth for their advice regarding the adjustment of the NBODY6 software package for our study. L.W. and X.C.Z. were supported by the Department of Astronomy at Peking University and the Kavli Institute for Astronomy and Astrophysics.
M.B.N.K. was supported by the Peter and Patricia Gruber Foundation through the PPGF fellowship, by the Peking University One Hundred Talent Fund (985), and by the National Natural Science Foundation of China (grants 11010237, 11050110414, 11173004). This publication was made possible through the support of a grant from the John Templeton Foundation and National Astronomical Observatories of Chinese Academy of Sciences. The opinions expressed in this publication are those of the author(s) do not necessarily reflect the views of the John Templeton Foundation or National Astronomical Observatories of Chinese Academy of Sciences. The funds from John Templeton Foundation were awarded in a grant to The University of Chicago which also managed the program in conjunction with National Astronomical Observatories, Chinese Academy of Sciences.  
R.P.C. was supported by the Swedish Research Council (grants 2012-2254 and 2012-5807). M.B.D. was supported by the Swedish Research Council (grants 2008-4089 and 2011-3991).



\begin{thebibliography}{999}

\bibitem[Aarseth(1999)]{aarseth1999} Aarseth, S.~J.\ 1999, \pasp, 111, 1333 

\bibitem[Aarseth(2003)]{aarseth2003} Aarseth, S.~J.\ 2003, Gravitational N-Body Simulations, Cambridge University, 2003

\bibitem[Abe et al.(2004)]{abe2004} Abe, F., Bennett, D.~P., 
Bond, I.~A., et al.\ 2004, Science, 305, 1264 

\bibitem[Adams et al.(2013)]{adams2013} Adams, F.~C., Anderson, 
K.~R., \& Bloch, A.~M.\ 2013, \mnras, 432, 438 

\bibitem[Allison et al.(2009a)]{allison2009a} Allison, R.~J., 
Goodwin, S.~P., Parker, R.~J., et al.\ 2009, \mnras, 395, 1449 

\bibitem[Allison et al.(2009b)]{allison2009b} Allison, R.~J., 
Goodwin, S.~P., Parker, R.~J., et al.\ 2009, \apjl, 700, L99 

\bibitem[Baumgardt \& Makino(2003)]{baumgardt2003} Baumgardt, H., \& Makino, J.\ 2003, \mnras, 340, 227 

\bibitem[Beaulieu et al.(2006)]{beaulieu2006} Beaulieu, J.-P., 
Bennett, D.~P., Fouqu{\'e}, P., et al.\ 2006, \nat, 439, 437 

\bibitem[Bihain et al.(2009)]{bihain2009} Bihain, G., Rebolo, R., Zapatero Osorio, M.~R., et al.\ 2009, \aap, 506, 1169 

\bibitem[Binney \& Tremaine(1987)]{binneytremaine} Binney, J., \& Tremaine, S.\ 1987, Princeton, NJ, Princeton University Press, 1987

\bibitem[Boley et al.(2012)]{boley2012} Boley, A.~C., Payne, 
M.~J., \& Ford, E.~B.\ 2012, \apj, 754, 57 

\bibitem[Caballero et al.(2006)]{caballero2006} Caballero, J.~A., Mart{\'{\i}}n, E.~L., Dobbie, P.~D., \& Barrado Y Navascu{\'e}s, D.\ 2006, \aap, 460, 635 

\bibitem[Caballero et al.(2007)]{caballero2007} Caballero, J.~A., B{\'e}jar, V.~J.~S., Rebolo, R., et al.\ 2007, \aap, 470, 903 

\bibitem[Cai et al.(2015)]{cai2015} Cai, M.~X., Spurzem, R., 
\& Kouwenhoven, M.~B.~N.\ 2015, arXiv:1501.01709 

\bibitem[Cervi{\~n}o et 
al.(2013a)]{cervino2013a} Cervi{\~n}o, M., Rom{\'a}n-Z{\'u}{\~n}iga, C., Luridiana, V., et al.\ 2013, \aap, 553, AA31 

\bibitem[Cervi{\~n}o et 
al.(2013b)]{cervino2013b} Cervi{\~n}o, M., Rom{\'a}n-Z{\'u}{\~n}iga, C., Bayo, A., et al.\ 2013, \aap, 553, AA32 

\bibitem[Chatterjee et al.(2008)]{Chatterjee2008} Chatterjee, S., 
Ford, E.~B., Matsumura, S., \& Rasio, F.~A.\ 2008, \apj, 686, 580 

\bibitem[Chatterjee et al.(2012)]{Chatterjee2012} Chatterjee, S., 
Ford, E.~B., Geller, A.~M., \& Rasio, F.~A.\ 2012, \mnras, 427, 1587 


\bibitem[Craig \& Krumholz(2013)]{craig2013} Craig, J., \& Krumholz, M.~R.\ 2013, \apj, 769, 150 

\bibitem[de Grijs et al.(2013)]{grijs2013} de Grijs, R., Li, C., 
Zheng, Y., et al.\ 2013, \apj, 765, 4 

\bibitem[de Juan Ovelar et 
al.(2012)]{dejuan2012} de Juan Ovelar, M., Kruijssen, J.~M.~D., Bressert, E., et al.\ 2012, \aap, 546, L1 

\bibitem[Delorme et al.(2012)]{delorme2012} Delorme, P., Gagn{\'e}, J., Malo, L., et al.\ 2012, \aap, 548, A26 

\bibitem[Di Stefano(2012)]{distefano2012} Di Stefano, R.\ 2012, 
\apjs, 201, 20 

\bibitem[Eggleton et al.(1989)]{eggleton1989} Eggleton, P.~P., Tout, 
C.~A., \& Fitchett, M.~J.\ 1989, \apj, 347, 998 

\bibitem[Eggleton et al.(1990)]{eggleton1990} Eggleton, P.~P., 
Fitchett, M.~J., \& Tout, C.~A.\ 1990, \apj, 354, 387 

\bibitem[Gahm et al.(2007)]{gahm2007} Gahm, G.~F., Grenman, T., 
Fredriksson, S., \& Kristen, H.\ 2007, \aj, 133, 1795  

\bibitem[Gaudi(2012)]{gaudi2012} Gaudi, B.~S.\ 2012, \araa, 50, 411 

\bibitem[Gould \& Loeb(1992)]{gould1992} Gould, A., \& Loeb, A.\ 1992, \apj, 396, 104 

\bibitem[Hao et al.(2013)]{hao2013} Hao, W., Kouwenhoven, 
M.~B.~N., \& Spurzem, R.\ 2013, \mnras, 433, 867 

\bibitem[Heggie(1975)]{heggie1975} Heggie, D.~C.\ 1975, \mnras, 
173, 729 

\bibitem[Heggie \& Hut(2003)]{heggiehut} Heggie, D., \& Hut, P.\ 2003, The Gravitational Million-Body Problem: A Multidisciplinary Approach to Star Cluster Dynamics, Cambridge University Press, 2003

\bibitem[Hurley et al.(2000)]{hurley2000} Hurley, J.~R., Pols, 
O.~R., \& Tout, C.~A.\ 2000, \mnras, 315, 543 

\bibitem[Hurley \& Shara(2002)]{hurley2002} Hurley, J.~R., \& Shara, M.~M.\ 2002, \apj, 565, 1251 

\bibitem[Juri{\'c} \& Tremaine(2008)]{Juric2008} Juri{\'c}, M., \& Tremaine, S.\ 2008, \apj, 686, 603 

\bibitem[Kaib et al.(2013)]{kaib2013} Kaib, N.~A., Raymond, 
S.~N., \& Duncan, M.\ 2013, \nat, 493, 381 

\bibitem[Kouwenhoven et al.(2005)]{kouwenhoven2005} Kouwenhoven, M.~B.~N., Brown, A.~G.~A., Zinnecker, H., Kaper, L., \& Portegies Zwart, S.~F.\ 2005, \aap, 430, 137 

\bibitem[Kouwenhoven(2006)]{kouwenhoven2006} Kouwenhoven, M.~B.~N.\ 
2006, Ph.D.~Thesis, University of Amsterdam 

\bibitem[Kouwenhoven et al.(2007)]{kouwenhoven2007} Kouwenhoven, M.~B.~N., Brown, A.~G.~A., Portegies Zwart, S.~F., \& Kaper, L.\ 2007, \aap, 474, 77 

\bibitem[Kouwenhoven et al.(2009)]{kouwenhovenpairing} Kouwenhoven, M.~B.~N., Brown, A.~G.~A., Goodwin, S.~P., Portegies Zwart, S.~F., \& Kaper, L.\ 2009, \aap, 493, 979 

\bibitem[Kouwenhoven et al.(2010)]{kouwenhoven2010} Kouwenhoven, 
M.~B.~N., Goodwin, S.~P., Parker, R.~J., et al.\ 2010, \mnras, 404, 1835 

\bibitem[Kouwenhoven et al.(2014)]{kouwenhoven2014} Kouwenhoven, 
M.~B.~N., Goodwin, S.~P., de Grijs, R., Rose, M., 
\& Kim, S.~S.\ 2014, \mnras, 445, 2256 

\bibitem[Kroupa et al.(1993)]{kroupatoutgilmore} Kroupa, P., Tout, C.~A., 
\& Gilmore, G.\ 1993, \mnras, 262, 545 

\bibitem[Lada \& Lada(2003)]{lada2003} Lada, C.~J., \& Lada, E.~A.\ 2003, \araa, 41, 57 

\bibitem[Liu et al.(2013)]{liu2013} Liu, H.-G., Zhang, H., 
\& Zhou, J.-L.\ 2013, \apj, 772, 142 

\bibitem[Lamers et al.(2005)]{lamers2005} Lamers, H.~J.~G.~L.~M., Gieles, M., Bastian, N., et al.\ 2005, \aap, 441, 117 

\bibitem[Luhman(2012)]{luhman2012} Luhman, K.~L.\ 2012, \araa, 50, 65 

\bibitem[Lucas \& Roche(2000)]{lucas2000} Lucas, P.~W., \& Roche, P.~F.\ 2000, \mnras, 314, 858 

\bibitem[Lucas et al.(2006)]{lucas2006} Lucas, P.~W., Weights, 
D.~J., Roche, P.~F., \& Riddick, F.~C.\ 2006, \mnras, 373, L60 

\bibitem[Malmberg et al. (2007)]{malmberg2007_encounters} Malmberg, D., de 
Angeli, F., Davies, M.~B., Church, R.~P., Mackey, D., 
\& Wilkinson, M.~I.\ 2007, \mnras, 378, 1207 

\bibitem[Malmberg et al. (2007)]{malmberg2007_kozai} Malmberg, D., Davies, 
M.~B., \& Chambers, J.~E.\ 2007, \mnras, 377, L1 

\bibitem[Malmberg et al.(2011)]{malmberg2011} Malmberg, D., Davies, 
M.~B., \& Heggie, D.~C.\ 2011, \mnras, 411, 859 

\bibitem[Mao \& Paczynski(1991)]{mao1991} Mao, S., \& Paczynski, B.\ 1991, \apjl, 374, L37 

\bibitem[Marsh et al.(2010)]{marsh2010} Marsh, K.~A., 
Kirkpatrick, J.~D., \& Plavchan, P.\ 2010, \apjl, 709, L158 

\bibitem[Moeckel \& Clarke(2011)]{moeckel2011} Moeckel, N., \& Clarke, C.~J.\ 2011, \mnras, 415, 1179 

\bibitem[Nagasawa \& Ida(2011)]{Nagasawa2011} Nagasawa, M., \& Ida, S.\ 2011, \apj, 742, 72 

\bibitem[Nowak et al.(2013)]{nowak2013} Nowak, G., Niedzielski, 
A., Wolszczan, A., Adam{\'o}w, M., \& Maciejewski, G.\ 2013, \apj, 770, 53 

\bibitem[Parker et al. (2009)]{parker2009_kozai} Parker, R.~J., \& Goodwin, S.~P.\ 2009, \mnras, 397, 1041 

\bibitem[Parker et al.(2009)]{parker2009_binaries} Parker, R.~J., Goodwin, 
S.~P., Kroupa, P., \& Kouwenhoven, M.~B.~N.\ 2009, \mnras, 397, 1577 

\bibitem[Parker \& Quanz(2012)]{parker2012} Parker, R.~J., \& Quanz, S.~P.\ 2012, \mnras, 419, 2448 

\bibitem[Pelupessy et 
al.(2013)]{pelupessy2013} Pelupessy, F.~I., van Elteren, A., de Vries, N., et al.\ 2013, \aap, 557, AA84 

\bibitem[Pe{\~n}a Ram{\'{\i}}rez et 
al.(2011)]{penaramirez2011} Pe{\~n}a Ram{\'{\i}}rez, K., Zapatero Osorio, M.~R., B{\'e}jar, V.~J.~S., Rebolo, R., \& Bihain, G.\ 2011, \aap, 532, A42 

\bibitem[Perets \& Kouwenhoven(2012)]{peretskouwenhoven} Perets, H.~B., \& Kouwenhoven, M.~B.~N.\ 2012, \apj, 750, 83 

\bibitem[Plummer (1911)]{plummer1911} Plummer, H.~C.\ 1911, \mnras, 
71, 460 

\bibitem[Portegies Zwart et al.(2010)]{portegies2010} Portegies Zwart, S.~F., McMillan, S.~L.~W., \& Gieles, M.\ 2010, \araa, 48, 431 

\bibitem[Portegies Zwart et al.(2013)]{portegies2013} Portegies 
Zwart, S., McMillan, S.~L.~W., van Elteren, E., Pelupessy, I., 
\& de Vries, N.\ 2013, Computer Physics Communications, 183, 456 

\bibitem[Poveda \& Allen(2004)]{poveda2004} Poveda, A., \& Allen, C.\ 2004, Revista Mexicana de Astronomia y Astrofisica Conference Series, 21, 49 

\bibitem[Quanz et al.(2010)]{quanz2010} Quanz, S.~P., Goldman, 
B., Henning, T., et al.\ 2010, \apj, 708, 770 

\bibitem[Rasio \& Ford(1996)]{Rasio1996} Rasio, F.~A., \& Ford, E.~B.\ 1996, Science, 274, 954 

\bibitem[Scholz et al.(2012)]{scholz2012} Scholz, A., 
Jayawardhana, R., Muzic, K., et al.\ 2012, \apj, 756, 24 

\bibitem[Spitzer(1987)]{spitzer1987} Spitzer, L.\ 1987, Princeton, 
NJ, Princeton University Press, 1987, 191 p.,  

\bibitem[Spurzem et al.(2009)]{spurzem2009} Spurzem, R., Giersz, 
M., Heggie, D.~C., \& Lin, D.~N.~C.\ 2009, \apj, 697, 458 

\bibitem[Strigari et al.(2012)]{strigari2012} Strigari, L.~E., 
Barnab{\`e}, M., Marshall, P.~J., 
\& Blandford, R.~D.\ 2012, \mnras, 423, 1856 

\bibitem[Sumi et al.(2011)]{sumi2011} Sumi, T., Kamiya, K., 
Bennett, D.~P., et al.\ 2011, \nat, 473, 349 

\bibitem[van Albada(1968)]{vanalbada1968} van Albada, T.~S.\ 1968, 
\bain, 20, 47 

\bibitem[Veras et al.(2011)]{veras2011} Veras, D., Wyatt, M.~C., 
Mustill, A.~J., Bonsor, A., \& Eldridge, J.~J.\ 2011, \mnras, 417, 2104 

\bibitem[Veras \& Tout(2012)]{veras2012} Veras, D., \& Tout, C.~A.\ 2012, \mnras, 422, 1648 

\bibitem[Veras \& Raymond(2012)]{verasscattering2012} Veras, D., \& Raymond, S.~N.\ 2012, \mnras, 421, L117 

\bibitem[Vereshchagin et al.(1987)]{vereshchagin1987} Vereshchagin, 
S.~V., Kraicheva, Z.~T., Popova, E.~I., Tutukov, A.~V., 
\& Yungelson, L.~R.\ 1987, Soviet Astronomy Letters, 13, 26 

\bibitem[Voyatzis et al.(2013)]{voyatzis2013} Voyatzis, G., 
Hadjidemetriou, J.~D., Veras, D., \& Varvoglis, H.\ 2013, \mnras, 430, 3383 

\bibitem[Weidner 
\& Kroupa(2004)]{weidner2004} Weidner, C., \& Kroupa, P.\ 2004, \mnras, 348, 187 

\bibitem[Weidner et al.(2013)]{weidner2013} Weidner, C., Kroupa, 
P., \& Pflamm-Altenburg, J.\ 2013, \mnras, 434, 84 

\bibitem[Zapatero Osorio et al.(2000)]{zapatero2000} Zapatero 
Osorio, M.~R., B{\'e}jar, V.~J.~S., Mart{\'{\i}}n, E.~L., et al.\ 2000, 
Science, 290, 103 



\end{thebibliography}
\end{document}